\documentclass[10pt,pre,aps,twocolumn,nofootinbib,tightenlines,superscriptaddress,amsmath,amssymb,final,letterpaper]{revtex4-1}
\usepackage[colorlinks=true,allcolors=blue]{hyperref}
\usepackage{amsmath,amssymb,graphicx}
\usepackage[shortlabels]{enumitem}
\usepackage{dsfont}
\newcommand{\eq}[1]{\begin{equation}\begin{aligned}#1\end{aligned}\end{equation}}
\usepackage{xcolor}
\usepackage{setspace}

\begin{document}

\title{Growing unlabeled networks} 

\author{Harrison Hartle}\thanks{\href{mailto:hhartle@santafe.edu}{hhartle@santafe.edu}}
\affiliation{Santa Fe Institute, Santa Fe, New Mexico 87505, USA}
\affiliation{Network Science Institute, Northeastern University, Boston, Massachusetts 02115, USA}

\author{Brennan Klein}
\affiliation{Network Science Institute, Northeastern University, Boston, Massachusetts 02115, USA}
\affiliation{Department of Physics, Northeastern University, Boston, Massachusetts 02115, USA}

\author{Dmitri Krioukov}
\affiliation{Network Science Institute, Northeastern University, Boston, Massachusetts 02115, USA}
\affiliation{Department of Physics, Northeastern University, Boston, Massachusetts 02115, USA}
\affiliation{Department of Mathematics, Northeastern University, Boston, Massachusetts 02115, USA}
\affiliation{Department of Electrical \& Computer Engineering, Northeastern University, Boston, Massachusetts 02115, USA}

\author{P. L. Krapivsky}
\affiliation{Santa Fe Institute, Santa Fe, New Mexico 87505, USA}
\affiliation{Department of Physics, Boston University, Boston, Massachusetts 02215, USA}

\date{\today}

\begin{abstract}
Models of growing networks are a central topic in network science. In these models, vertices are usually labeled by their arrival time, distinguishing even those node pairs whose structural roles are identical. In contrast, unlabeled networks encode only structure, so unlabeled growth rules must be defined in terms of structurally distinguishable outcomes; network symmetries therefore play a key role in unlabeled growth dynamics. Here, we introduce and study models of growing unlabeled trees, defined in analogy to widely-studied labeled growth models such as uniform and preferential attachment. We develop a theoretical formalism to analyze these trees via tracking their leaf-based statistics. We find that while many characteristics of labeled network growth are retained, numerous critical differences arise, caused primarily by symmetries among leaves in common neighborhoods. In particular, degree heterogeneity is enhanced, with the strength of this enhancement depending on details of growth dynamics: mild enhancement for uniform attachment, and extreme enhancement for preferential attachment. These results and the developed analytical formalism may be of interest beyond the setting of growing unlabeled trees.
\end{abstract}

\maketitle

\section{Introduction}\label{sec:introduction}

Probabilistic models of network growth are ubiquitous in network science \cite{barabasi2013network, newman2018networks, van2014random} and have attracted wide attention with numerous applications to real growing networks \cite{kunegis2013preferential, wang2008measuring, jeong2003measuring, capocci2006preferential, akbacs2015preferential, piva2021networks, maoz2012preferential}. Growth mechanisms include uniformly random attachment \cite{krapivsky2010kinetic}, preferential attachment based on structural properties of nodes such as degree \cite{Simon,barabasi1999emergence, krapivsky2001organization}, influences of node-variables such as vertex-intrinsic fitness \cite{bianconi2001competition}, spatial geometry \cite{zuev2015emergence, papadopoulos2012popularity}, and community assignment \cite{shang2020growing}, among numerous other variations \cite{moore2006exact, dorogovtsev2002evolution, zheng2021scaling}. Network growth models have been formulated across a range of sample spaces including growing directed graphs \cite{karrer2009random}, growing weighted graphs \cite{PhysRevE.71.036124,PhysRevE.71.026103,barrat2004modeling}, growing simplicial complexes \cite{courtney2017weighted}, growing hypergraphs \cite{avin2019random, krapivsky2023random, roh2023growing}, and growing causal sets~\cite{rideout1999classical, surya2025causal}.
   
However, in the aforementioned studies, the underlying growing objects are usually {\it labeled}, with vertices indexed by arrival order and hence distinguishable even if occupying identical structural roles in the network. In contrast, {\it unlabeled} graphs represent network {\it structure} as a pattern of connectivity among otherwise indistinguishable actors. The choice of an unlabeled representation has statistical consequences, as previously shown in non-growing random unlabeled networks \cite{schwenk1977asymptotic, luczak1991deal, paton2022entropy, evnin2025ensemble}. Yet the manifestation of unlabeled statistics in network growth has not previously been explored. What are the unlabeled analogs to the most widely studied models of labeled network growth, and how does their phenomenology compare to that of their labeled counterparts? 

To address these questions, we formulate models of unlabeled tree growth and develop an analytical framework for studying them; we consider the unlabeled versions of uniform attachment (random recursive trees) and preferential attachment, including the Barab\'asi-Albert model and preferential attachment with an additive shift, i.e., initial attractiveness of vertices. Unlabeled growth rules are probability distributions over the set of unlabeled trees that could arise via a single leaf-attachment event; as such, attachment probabilities are not tilted towards structural positions with higher numbers of representatives. Instead, each meaningfully different (structurally distinct) growth outcome is assigned equal a priori probability. This leads to effects such as suppressed attachment to leaves, since all leaves of a common neighbor represent the same unlabeled attachment possibility. As a result, unlabeled growth models exhibit enhanced degree-heterogeneity of varying severity depending on the specific attachment rule. We find a geometric degree distribution with base $c\approx 0.57$ in unlabeled uniform attachment, in contrast with the labeled model's value of $\frac{1}{2}$; in unlabeled preferential attachment, we find an extremely heavy-tailed degree distribution with powerlaw exponent $\gamma_0\approx 1.84$ alongside anomalous scaling and leaf-proliferation phenomena, in contrast with the labeled model's exponent $3$ and linear scaling. We also show that unlabeled preferential attachment with an additive shift leads to powerlaw degree distributions with exponent value $\gamma\in(1,\infty)$, notably including the anomalous scaling regime $\gamma\in(1,2]$, in contrast with the range of tail exponents $(2,\infty)$ accessible to labeled preferential attachment with additive shift. We postpone further discussion of our results and potential extensions to Sec.~\ref{sec:discussion}.

\begin{figure}[t]
    \centering
    \includegraphics[width=1.3\linewidth, trim = 1400 500 900 500, clip]{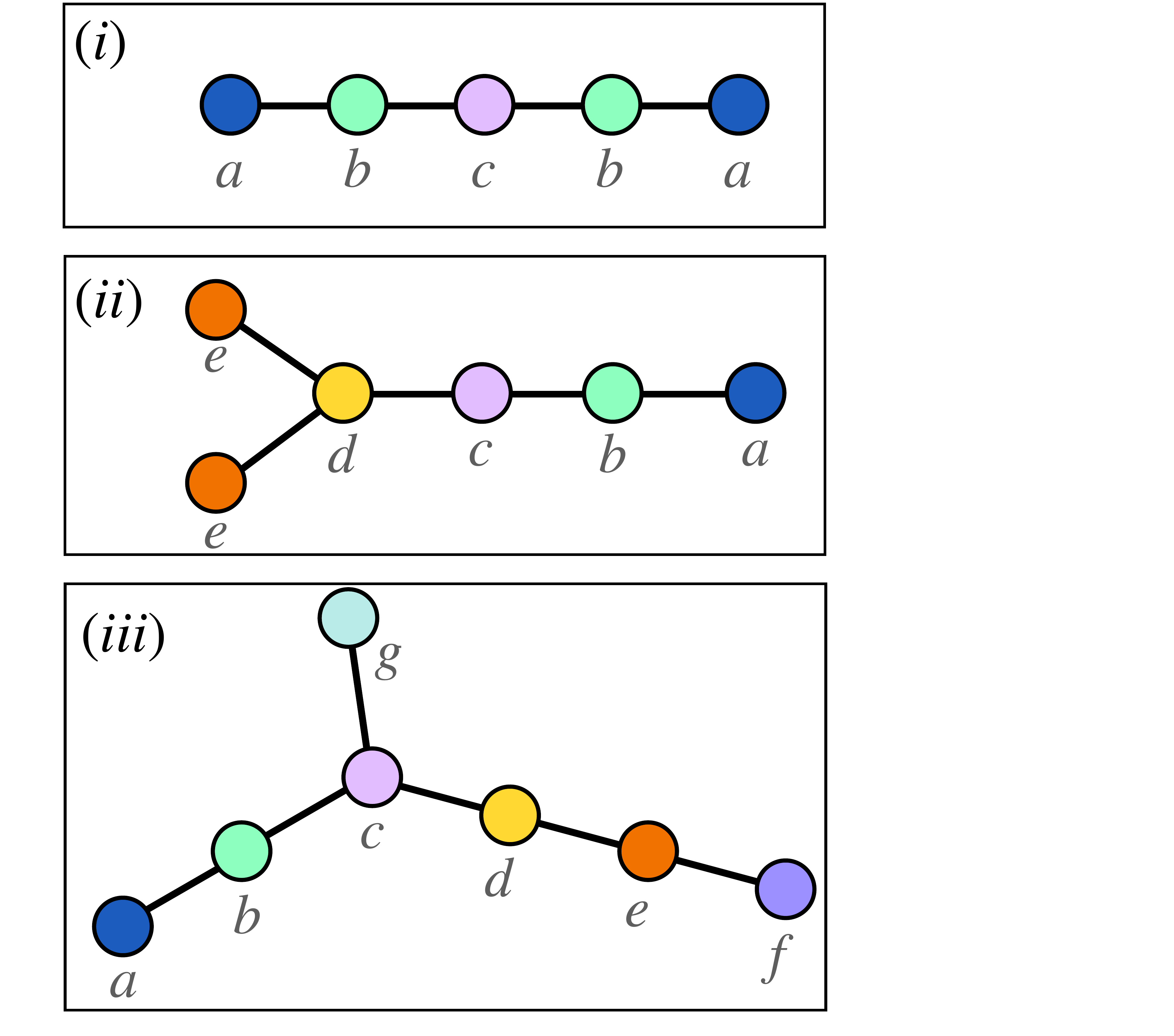}
    \caption{ Symmetries and orbits in tree graphs. (i) a tree with nonlocal and non-leaf symmetries: the orbits $a$ and $b$; (ii) a leaf-symmetric tree (orbit $e$ consists of two leaves); (iii) an asymmetric tree (all orbits of unit size); Nodes are colored according to structural position (automorphism orbit).}
    \label{fig:symmetry_schematic}
\end{figure}

We organize the presentation as follows. In Sec.~\ref{sec:formulation} we formulate growth rules for unlabeled networks, obtaining unlabeled analogs to popular labeled growth models. In Sec.~\ref{sec:leaf_based} we develop an analytical framework to study them. In Sec.~\ref{sec:results} we apply this framework to analyze the properties of networks from Sec.~\ref{sec:formulation}, comparing them to the corresponding results for labeled growth. In Sec.~\ref{sec:discussion} we describe outstanding questions, challenges, and implications.

\section{Unlabeled growth models}\label{sec:formulation}

In this section, we formulate the notion of unlabeled growth rules (Sec.~\ref{ssec:growth_rules}), then define several models of unlabeled growth (Sec.~\ref{ssec:growing_unlabeled}) in analogy to the well-studied labeled growth models: linear preferential attachment (PA) \cite{krapivsky2000connectivity, dorogovtsev2000structure}, and as special cases the Barab\'asi-Albert (BA) model \cite{barabasi1999emergence} and the random recursive tree (RRT) \cite{meir1974cutting}. For background on (nongrowing) unlabeled random graphs, see Appendix~\ref{app:random_unlabeled}; for background on (labeled) network growth models, see Appendix~\ref{app:labeled_growth}.

\subsection{Unlabeled growth rules}\label{ssec:growth_rules}

Unlabeled growth statistics are as follows: first, the set of distinct attachment options is evaluated; second, a probability distribution over those attachment options is specified---by default, uniformly---but with arbitrary attachment kernels atop that; third, a growth step is sampled from those options, yielding a tree of size one larger. A tree growth trajectory is denoted $(G_3,...,G_n) \in\mathcal{G}_3\times \cdots \times \mathcal{G}_n$, with $\mathcal{G}_n$ denoting the set of $n$-node trees growable by a sequence of leaf-attachments. The initial size is $3$ as the trajectory $G_1,G_2,G_3$ is determined structurally (an isolate, a dyad, then a two-leaf star, i.e., $3$-node line graph). We consider a Markovian dependence, with growth steps governed by a sequence of conditional distributions $P_n(G_n|G_{n-1})$. The set of possible graphs $G'$ arising from a single attachment to a vertex in $G$ is denoted $\mathcal{G}(G)$. In the labeled setting, $\mathcal{G}(G)$ is in one-to-one correspondence with the set of labeled vertices $V(G)$. In the unlabeled setting, $\mathcal{G}(G)$ corresponds to the set of distinct structural positions (orbits) $u\in D(G)$. (This direct correspondence between attachment possibilities and orbits is for unlabeled {\it trees}; see Appendix~\ref{app:non_tree} for discussion of non-tree growth.)

Each such structural position $u\in D(G)$ is a group of structurally indistinguishable nodes. For instance, positions $a,b$ in Fig.~\ref{fig:symmetry_schematic}(i), and position $e$ in Fig.~\ref{fig:symmetry_schematic}(ii), are structural positions with two representatives each. Often, nodes are the unique representatives of a given structural position (e.g., position $c$ in Fig.~\ref{fig:symmetry_schematic}(i), and each of $d,c,b,a$ in Fig.~\ref{fig:symmetry_schematic}(ii)). Graphs for which {\it all} nodes have distinct positions are called {\it asymmetric}; for example Fig.~\ref{fig:symmetry_schematic}(iii). Groups of equivalent nodes are called {\it orbits}, short for {\it automorphism orbits} in labeled graph contexts \cite{doi:https://doi.org/10.1002/9783527627981.ch1}.

Crucially, for the setting of {\it growing} trees, attachment by a new node to any member of a group of structurally identical nodes leads to the {\it same} unlabeled tree. An unlabeled tree growth rule $P_n(G'|G)$ is thus equivalent to a distribution over orbits $u\in D(G)$. The probability of a new leaf attaching to a node of orbit $u$ is denoted
\eq{
p_u:=\mathbb{P}(v(G,G')=u),
}
where $v(G,G')\in D(G)$ is the orbit of $G$ attached into to obtain $G'\in\mathcal{G}(G)$. This distribution over orbits replaces the distribution over structure and choice of labeled representative, in labeled growth models. If $G_u$ denotes the graph in $\mathcal{G}(G)$ obtained by attaching to orbit $u$, then $\mathcal{G}(G)=\{G_u\}_{u\in D(G)}$.

We consider growth rules of the form
\begin{equation}
\label{eq:attachment_kernel}
P_n(G'|G)=\frac{f(G,G')}{\sum_{G''\in\mathcal{G}(G)}f(G,G'')},
\end{equation}
where $f$ is a real-valued function called the {\it attachment kernel}. Degree-based attachment schemes, among other structure-informed attachment rules, are functions of {\it node invariants} \cite{fortin1996graph}: node properties common to all representatives of a given orbit (i.e., properties independent of node label and any node-associated variables; examples include degree, clustering coefficient, betweenness centrality, average nearest neighbor degree, etc.). 

Any labeled attachment rule based on node invariants can also be formulated for unlabeled growth. Namely, $f(G,G')=\phi(G,v(G,G'))$ for the chosen node invariant $\phi(G,u)\ge 0$, a function of orbit $u\in D(G)$. In the labeled case, this implies that all nodes in a given orbit have an equal probability of being attached to. The unlabeled growth rule is expressed in terms of the node invariant $\phi$ as
\eq{
p_u=\frac{\phi(G,u)}{\sum_{u'\in D(G)}\phi(G,u')}.
}

The node invariants considered herein for growth rules are {\it degree}-based. In particular, we consider $\phi(G,u)=k_u(G)+\delta$ for $\delta\ge -1$, with $k_u(G)$ the degree of nodes in orbit $u\in D(G)$; see Sec.~\ref{sssec:upa}. Furthermore, we rely on much more information-rich {\it complete} node invariants for statistically exact algorithmic generation of unlabeled growth (see Appendix~\ref{app:simulations}).

\begin{figure}[t]
    \centering
    \includegraphics[width=1.3\linewidth, trim = 1350 450 900 500, clip]{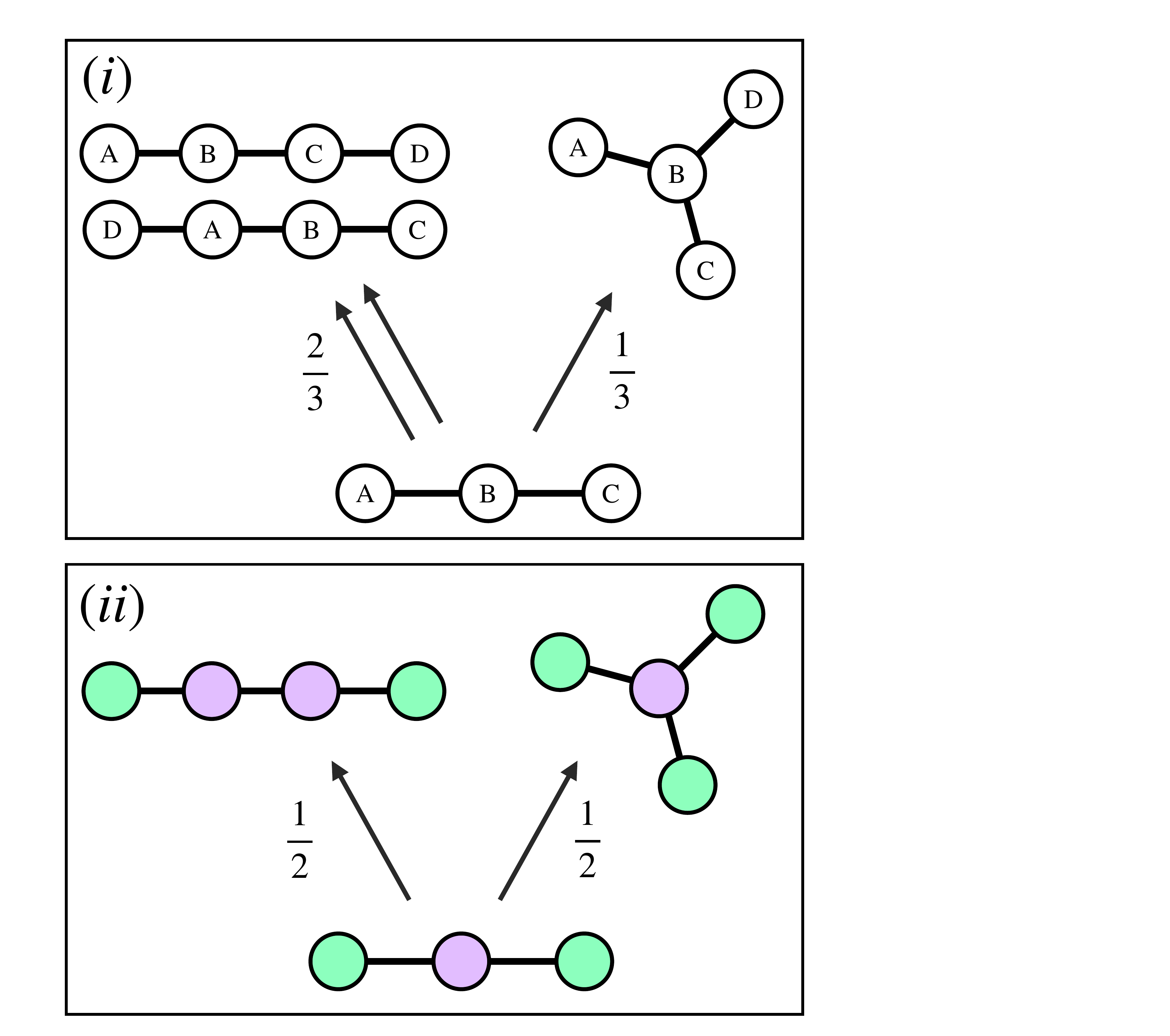}
    \caption{Labeled vs unlabeled tree growth. (i) a labeled growth step, with $2/3$ probability of producing a line graph and $1/3$ probability of producing a star graph; (ii) an unlabeled growth step, with balanced probabilities $(1/2,1/2)$ of producing a line graph of star graph. Nodes in (i) are labeled alphabetically, and in (ii) are colored according to structural position (orbit).}
    \label{fig:growth_schematic}
\end{figure}

\subsection{Models of growing unlabeled trees}\label{ssec:growing_unlabeled}

The simplest example of unlabeled growth is unlabeled uniform attachment (UUA). For an illustration of how it differs from the labeled RRT, see Fig.~\ref{fig:growth_schematic}. The unbiased {\it labeled} growth process results in probabilities $(\frac{1}{3},\frac{2}{3})$ of the star graph and line graph at size $n=4$, respectively---a nonuniform distribution over the two shapes $(S_4,L_4)$. In contrast, the {\it unlabeled} RRT assigns equal probabilities $(\frac{1}{2},\frac{1}{2})$ to the star and the line---a fair coin, or unbiased model, from the unlabeled standpoint. 

In what follows, we formulate unlabeled growth models in direct analogy to their well-studied labeled counterparts by using the same form of attachment kernel $f(G,G')$, subsequently determining the growth rule by Eq.~\eqref{eq:attachment_kernel}, to result in unlabeled uniform attachment (Sec.~\ref{sssec:uua}), unlabeled preferential attachment (Sec.~\ref{sssec:upa0}), and the more general unlabeled preferential attachment with initial attractiveness (Sec.~\ref{sssec:upa}). Recall that in all of these models, for a given unlabeled tree $G$, the set of possible growth outcomes $\mathcal{G}(G)$ corresponds to attachments to $G$'s {\it orbits} $D(G)$. Orbits represent distinct structural positions of vertices in the tree $G$, corresponding in turn to distinct unlabeled outcomes of an attachment. Note, $|D(G)|\le |V(G)|$, with equality if $G$ is asymmetric \cite{erdos1963asymmetric} (akin to having only the identity as an automorphism \cite{mckay1984automorphisms}, in the labeled setting). We herein consider the {\it degree} $k_u$ as the node invariant of interest, examining kernels that are constant or linearly varying with $k_u$. See Table~\ref{tab:models} for the models and their parametric relationships.

\subsubsection{Unlabeled uniform attachment}\label{sssec:uua}

The unlabeled uniform attachment ($\mathrm{UUA}$) model is the unlabeled version of a random recursive tree; Fig.~\ref{fig:growth_schematic}(i),(ii) which compare labeled and unlabeled uniform attachment for the growth step from $n=3$ to $n=4$. $\mathrm{UUA}$ is the most simple and unbiased unlabeled growth rule, with attachment kernel being a constant, $f(G,G')=1$ for all $G'\in\mathcal{G}(G)$. Equivalently, $\phi(G,u)=1$ for all $u\in D(G)$. This lack of attachment preference results in a uniform distribution across the set of structurally distinct trees arising from possible attachments to $G$. Mathematically, the unlabeled growth rule is equivalent to a uniform distribution across orbits, hence having probability value
\eq{
p_u^{\mathrm{UUA}}=\frac{1}{|D(G)|}
}
for each $u\in D(G)$. In this model, the degree distribution appears to have geometric tail $p_k\sim c^k$, where $c\approx 0.57$ is the asymptotic fraction of leaves; we obtain $c$ analytically and analyze leaf-related statistics using techniques developed below in Sec.~\ref{sec:leaf_based}; see Fig.~\ref{fig:uua_leafdeg}.

\subsubsection{Unlabeled preferential attachment}\label{sssec:upa0}

The {\it preferential attachment} (PA) tree \cite{Simon,barabasi1999emergence,price1976general} is a widely studied mechanistic model of cumulative advantage. It exhibits a scale-free degree distribution $p_k=4/(k(k+1)(k+2))$, having unbounded second moment $\langle k^2\rangle\sim \log n$. The growth rule incorporates direct degree-proportionality, which in the unlabeled form is expressed as $f(G,G')=k_{u(G,G')}$, where $u(G,G')\in D(G)$ denotes the orbit attached to in $G$ to produce $G'$, and where $k_u$ denotes the degree of all nodes in orbit $u\in D(G)$; equivalently, $\phi(G,u)=k_u$. Thus the unlabeled preferential attachment (UPA) model has growth rule
\eq{
p_u^{\mathrm{UPA}}=\frac{k_{u}}{\sum_{u'\in D(G)}k_{u'}}.
}

As we show in Sec.~\ref{sec:results}, this model exhibits scale-free {\it leaf-proliferation}, admitting super-hubs; $\langle\bar{k}_{\mathrm{max}}\rangle\sim n$. In particular, its degree distribution $p_k\sim k^{-\gamma_0}$ has anomalous tail exponent $\gamma_0<2$ (namely, $\gamma_0=1+\xi\approx 1.84$) among nonleaves, which constitute a vanishing fraction of nodes. This resembles the behavior of {\it isotropic redirection} \cite{krapivsky2017emergent}. We obtain $\xi$ analytically in Sec.~\ref{ssec:upa0_results} via methods developed below in Sec.~\ref{sec:leaf_based}; see also Appendix~\ref{app:exponent_finding}.

\subsubsection{Unlabeled preferential attachment with additive shift}\label{sssec:upa}

Unlabeled linear preferential attachment with additive shift, $\mathrm{UPA}(\delta)$, is defined by $f(G,G')=k_{v(G,G')}+\delta$, or equivalently $\phi(G,u)=k_u+\delta$, with some additive shift parameter $\delta\in [-1,\infty)$, with $v(G,G')\in D(G)$ the orbit attached into by an arriving leaf in order to obtain $G'\in\mathcal{G}(G)$, and with $k_u$ denoting the degree of any given orbit $u\in D(G)$. The growth rule for unlabeled linear preferential attachment ($\mathrm{UPA}(\delta)$), by Eq.~\eqref{eq:attachment_kernel}, is thus
\eq{
p_u^{\mathrm{UPA}(\delta)}=\frac{k_{u}+\delta}{\sum_{u\in D(G)}k_u+\delta|D(G)|},
}
for each $u\in D(G)$. The parameter $\delta\in[-1,\infty)$ is the {\it initial attractiveness} of vertices \cite{dorogovtsev2000structure}. At $\delta=-1$, all leaves are unable to be attached to, resulting in a star graph with probability~$p_n(S_n)=1$ for all $n$. At $\delta=\infty$, any influence of degree heterogeneity is suppressed, resulting in the $\mathrm{UUA}$ growth rule $p_u=1/|D(G)|$ (Sec.~\ref{sssec:uua}). At $\delta=0$, the kernel becomes $f(G,G')=k_{v(G,G')}$, resulting in the unlabeled BA model or $\mathrm{UPA}:=\mathrm{UPA}(0)$ (Sec.~\ref{sssec:upa0}).

We find that leaf-proliferation and the associated anomalous scaling and powerlaw tail exponent $\gamma(\delta)\le 2$ are exhibited for $\delta\in(0,\delta^*]$ with the threshold value $\delta^*\approx 0.19$ determined analytically. We also show linear scaling and $\gamma\ge 2$ for $\delta\ge \delta^*$, and obtain the asymptotic leaf-fraction as a function of $\delta$, approaching the $\mathrm{UUA}$ value $c\approx 0.57$ as $\delta\rightarrow\infty$. These and related results are presented in Sec.~\ref{ssec:upa_results}.

\subsection{Delabeled growth}\label{ssec:delabeled_growth}

In this section we further describe the relationship between labeled and unlabeled growth statistics, using the notion of \textit{delabeled} models.

We begin by addressing the unfamiliarity of probabilistic models of unlabeled objects. One might argue that new nodes ``should'' attach into an orbit with probability proportional to that orbit's number of representatives, because larger orbits contain more nodes and thus represent more ways to connect. However, in the unlabeled setting, the contrary is true; the nodes of a given orbit all represent one and the same attachment possibility: they all lead to the same unlabeled tree outcome. Hence, tipping the scales towards attachment to larger orbits constitutes a statistical bias (see Eq.~\eqref{eq:growth_bias}). That biased distribution over unlabeled outcomes is called a {\it delabeled} model \cite{luczak2017structural, paton2022entropy}. Delabeled models exhibit {\it labeled} statistics but viewed in a sample space of unlabeled objects. 

Unlabeled growth rules can recover delabeled statistics for growth rules based on node invariants. Suppose the labeled growth has attachment kernel $f_0(G,G')=\phi_0(G,u)$. An unlabeled growth model capturing the resulting delabeled statistics has attachment kernel 
\eq{
\label{eq:growth_bias}
f(G,G')=|v(G,G')|f_0(G,G'),
}
where $v(G,G')\in D(G)$ denotes the orbit attached to in $G$ to obtain $G'$, and $|v(G,G')|$ denotes the number of nodes therein. In terms of the node invariant $\phi_0$ defining the labeled growth rule, the delabeled growth rule can be represented via unlabeled growth with size-biased node invariant
\eq{
\label{eq:growth_bias2}
\phi(G,u)=|u|\phi_0(G,u).
}
Eqs.~\eqref{eq:growth_bias}~and~\eqref{eq:growth_bias2} express a label-induced bias in growth tendency analogous to that arising in delabeled static random graphs \cite{veitch2015class, caron2017sparse}. In particular, for {\it exchangeable} random labeled graphs \cite{diaconis2007graph} (whose probabilities are functions of {\it graph} invariants \cite{brigham1985compilation,randic1993search}), the probability $P_D(U)$ of delabeled outcome $U$ is
\eq{
P_D(U)=|\mathrm{Iso}(G_U)|P_L(G_U),
}
with $P_L(G)$ the probability of labeled graph $G$ and $G_U$ denoting any labeled version of $U$, and with $\mathrm{Iso}(G)$ the set of all labeled isomorphisms of $G$. The delabeled attachment kernel (Eq.~\eqref{eq:growth_bias}, Eq.~\eqref{eq:growth_bias2}) likewise describes a delabeling-induced bias. 

Delabeled growth models have appeared in several settings before, as a means of studying labeled growth processes; the entropy of delabeled preferential attachment has been characterized \cite{luczak2017structural}, and delabeled growth models are the problem setting for growth history reconstruction problems including root inference \cite{young2019phase, timar2020choosing} and patient zero estimation \cite{10.1111/rssb.12428}.

\begin{table*}
\begin{tabular}{|c|c|c|c|c|}
\hline
\textbf{Unlabeled growth model} & \textbf{Abbreviation} & \textbf{Labeled case}\\ 
\hline
Unlabeled shifted preferential attachment & $\mathrm{UPA}(\delta)$ & $\mathrm{PA}(\delta)$ \cite{krapivsky2000connectivity, dorogovtsev2000structure, van2014random}\\
\hline
Unlabeled preferential attachment & $\mathrm{UPA}$ & $\mathrm{PA}$/$\mathrm{BA}$ \cite{barabasi1999emergence} \\ 
\hline
Unlabeled uniform attachment & $\mathrm{UUA}$ & $\mathrm{UA}$/$\mathrm{RRT}$ \cite{smythe_mahmoud_1995, krapivsky2010kinetic} \\
\hline
\end{tabular}
\caption{Unlabeled growth models.}
\label{tab:models}
\end{table*}

\section{Leaf-based analytical framework}\label{sec:leaf_based}

The nature of unlabeled growth is analytically challenging in its full generality, due to symmetries unpredictably forming and breaking as networks grow. Here, we present a tractable analytical approach: a description in terms of leaf-based variables, namely, how many leaves exist in total and how many nonleaves have exactly $\ell$ leaf-neighbors, for each $\ell\ge 0$. These analytical techniques hold under the assumption of the network having only localized leaf-based symmetries; we find that this approximation holds in the settings of interest herein.

\subsection{Leaf-symmetric trees and leaf-statistic variables}\label{ssec:method_intro}

We describe a set of variables for which analytical traction is possible in the setting of unlabeled tree growth. The approach works for trees in which all symmetries are among {\it leaves}, specifically among each set of leaves with the same non-leaf neighbor. We call graphs in which symmetries are only among leaves in common neighborhoods {\it leaf-symmetric}. See Fig.~\ref{fig:symmetry_schematic}(i), (ii), and (iii) for an illustration comparing nonlocal symmetries, leaf-symmetries, and a lack of any symmetries.
 
Under leaf-symmetric conditions, an accounting of leaf-variable evolution sidesteps the possibility of intractable nonlocal symmetries. The necessary and sufficient condition for being leaf-symmetric is that the induced subgraph among all nonleaves, when colored by their numbers of leaf-neighbors, is an {\it asymmetric colored graph}. As it turns out, many growing trees of interest are leaf-symmetric, or at least approximately leaf-symmetric, enabling a consistent description via the variables ($L,\mathbb{M}$) introduced below. 

The {\it leaf-degree} $\ell_i$ of a non-leaf node $i$ is defined as the number of $i$'s neighbors that are leaves \cite{zhou2024d}. The leaf-degree $\ell_i$ and degree $k_i$ respectively satisfy $\ell_i\ge 0$, $k_i\ge 1$, and
\eq{
\frac{1}{2}\sum_{i}k_i&=n-1,\\
\sum_{i}\ell_i&=L,
}
with $n$ and $L$ denoting the total number of nodes and leaves, respectively. The number of nonleaves with leaf-degree $\ell$ ($\ell\ge 0$) is denoted $M_{\ell}$. Then $M_0$ is the number of non-leaf nodes without any leaf neighbors (``protected nodes'' \cite{MAHMOUD20122218, mahmoud2015asymptotic}). For a schematic visualization of the variables $\mathbb{M}:=\{M_{0},M_1,...\}$, see Fig.~\ref{fig:Ml_schematic}. The following key relations among $(n,L,\mathbb{M})$ hold:
\eq{
\label{eq:n_norm}
L+\sum_{\ell\ge 0}M_{\ell}=n,\\
\sum_{\ell\ge 0}lM_{\ell}=L.
}
Combining the above yields consistency condition $n=\sum_{\ell\ge 0 } (\ell+1)M_{\ell}$ between $n$ and $\mathbb{M}$. 

Recall, the empirical degree distribution is
\eq{
p_k=\frac{N_k}{n}, \ k\ge 0,
}
where $N_k$ denotes the number of nodes of degree $k$, so that $\sum_{k\ge 0}N_k=n$, i.e., $\sum_{k\ge 0}p_k=1$. In analogy, the {\it leaf-degree distribution} \cite{li2008gnutella} can be written as
\eq{
r_{\ell }=\frac{M_{\ell}}{n-L},
}
with denominator $\sum_{\ell\ge 0}M_{\ell}=n-L$, so that $\sum_{\ell\ge 0}r_{\ell }=1$; $r_{\ell }$ is the probability that a randomly selected non-leaf has leaf-degree $\ell$.

In general, a graph $G$ with orbits $D(G)$ has a number of symmetries equal to 
\eq{
|\mathrm{Aut}(G)|=\prod_{u\in D(G)}|u|!,
}
representing independent indistinguishability of nodes within each orbit. An asymmetric tree $G$ has $|\mathrm{Aut}(G)|=\prod_{i=1}^n1=1$, each permutation yielding a distinct isomorphic graph. A leaf-symmetric tree has number of symmetries expressible in terms of $\{M_{2},M_3,...\}$, namely,
\eq{
|\mathrm{Aut}(G)|=\prod_{\ell\ge 2}(\ell !)^{M_{\ell}}.
}

Drawing on these preliminaries, we walk through how growth steps occur in leaf-symmetric trees. The choice of probabilistic model sets the attachment probabilities, but the attachment outcomes are structurally determined (see Eq.~\eqref{eq:update_leafattachment}, Eq.~\eqref{eq:update_nonleafattachment}). Hence the following considerations apply for any Markovian tree growth model, and for both labeled and unlabeled growth---as long as the process favors the emergence of leaf-symmetric trees. We will later specialize to the cases of unlabeled uniform attachment (UUA) and preferential attachment (UPA).

\subsection{Description of unlabeled leaf-symmetric tree growth}\label{ssec:unlabeled_leaf_symm_growth}

Herein we introduce a description of how $(L,\mathbb{M})$ evolve under leaf-symmetric Markovian tree growth in terms of a stochastic update rule. A each size $n\ge 3$, the variables $(L,\mathbb{M})$ at size $n+1$ are determined according to a growth rule with outcome probabilities $(P_1,P_2,...,R_0,R_1,R_2,...)$, satisfying 
\eq{
\label{eq:normalization}
\sum_{\ell\ge 1}P_{\ell }+\sum_{\ell\ge 0}R_{\ell }=1,
}
where $P_{\ell }$ is the probability of attaching to a leaf with multiplicity $\ell$ ($\ell\ge 1$), and where $R_{\ell }$ is the probability of attaching to a leaf-degree-$\ell$ node ($\ell\ge 0$). The probabilities $(P_{\ell })_{\ell\ge 1}$, $(R_{\ell })_{\ell\ge 0}$ are functions of $(n,L,\mathbb{M})$ determined by the choice of growth model.

\begin{figure}[t]
    \centering
    \includegraphics[width=1.0\linewidth]{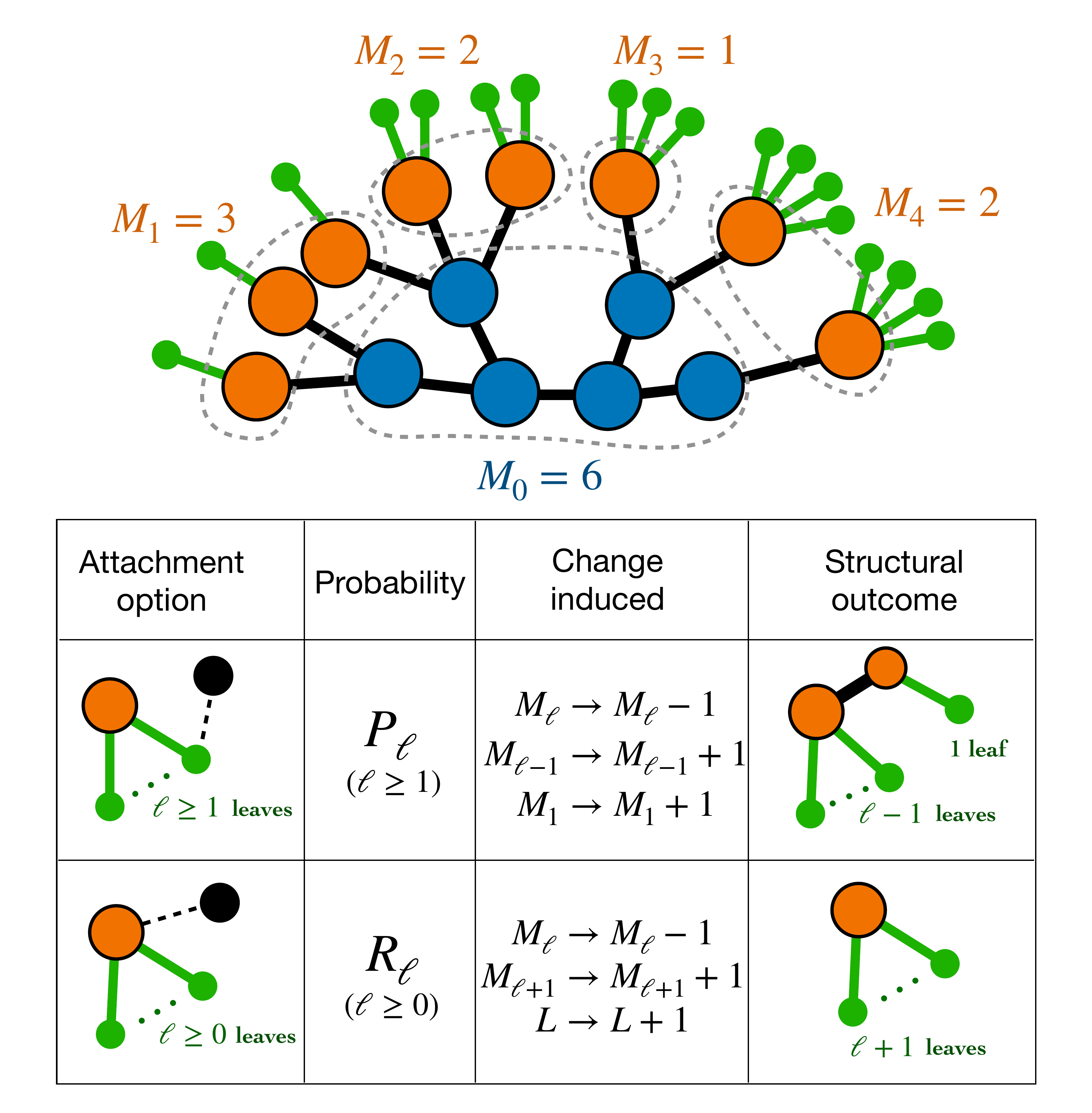}
    \caption{(Above) The variables $\mathbb{M}=\{M_{0},M_1,...\}$ count the numbers of nodes with exactly $\ell$ leaf-neighbors (leaf-degree $\ell$). (Below) The different types of growth steps (attach to non-leaf, attach to leaf), their probabilities, their changes induced in the variables $(L,\mathbb{M})$, and the associated structural outcomes.}
    \label{fig:Ml_schematic}
\end{figure}

Starting at $n=3$ with $L=2$ and $M_{\ell}=\delta_{\ell,2}$ (the 3-node star graph), the following is iterated at each growth step $n\rightarrow n+1$. A leaf may be attached to: with probability $P_{\ell }$ ($\ell\ge 1$), the variables update as
\eq{
\label{eq:update_leafattachment}
M_{\ell}&\rightarrow M_{\ell}-1, \\
M_{\ell - 1}&\rightarrow M_{\ell - 1}+1,\\ 
M_1&\rightarrow M_1+1, 
}
encoding a node of leaf-degree $\ell$ becoming a node of leaf-degree $\ell-1$, and that the leaf attached to becomes a node of leaf-degree one (see Fig.~\ref{fig:Ml_schematic}). Alternatively, a non-leaf may be attached to: with probability $R_{\ell }$, a non-leaf of leaf-degree $\ell\ge 0$ is attached to, in which case
\eq{
\label{eq:update_nonleafattachment}
M_{\ell}&\rightarrow M_{\ell}-1, \\ M_{\ell + 1}&\rightarrow M_{\ell + 1}+1, \\ 
L&\rightarrow L+1,
}
reflecting that a node with leaf-degree $\ell$ becomes a node with leaf-degree $\ell+1$, and that a new leaf is added (Fig.~\ref{fig:Ml_schematic}, lower row). (Note regarding Eq.~\eqref{eq:update_leafattachment}: in the case of $\ell=1$, $M_1\rightarrow M_1-1$ is canceled by the $M_1\rightarrow M_1+1$, leaving only the change $M_0\rightarrow M_0+1$.)

The leaf-statistic update rules (Eq.~\eqref{eq:update_leafattachment}, Eq.~\eqref{eq:update_nonleafattachment}), and the choices of $(P_{\ell })_{\ell\ge 1}$ and $(R_{\ell })_{\ell\ge 0}$ as a function of $(n,L,\mathbb{M})$ determine a stochastic evolution of leaf-statistic variables, amenable to exact simulation and analytical solution. 

The expected values of changes in leaf-statistics at step $n\rightarrow n+1$ can be expressed in terms of the variables at size $n$ by making use of Eq.~\eqref{eq:update_leafattachment} and  Eq.~\eqref{eq:update_nonleafattachment}. In particular,
\eq{
\label{eq:evol_general}
\dot{L}&=\sum_{\ell'\ge 0}R_{\ell' },\\
\dot{M}_0&=P_1-R_0,\\ 
\dot{M}_1&= P_2+R_0-R_1+\sum_{\ell'\ge 2}P_{\ell' }\\
\dot{M}_{\ell }&=P_{\ell + 1}-(P_{\ell }+R_{\ell })+R_{\ell - 1}, \ \  \ell\ge 2,
}
with notation $\dot{X}$ denoting the expected change in quantity $X$ at step $n\rightarrow n+1$. We assume the random variables $(L,M_0,M_1,...)$ are concentrated around their means. Eqs.~\eqref{eq:evol_general} are evaluated for unlabeled uniform attachment (UUA) in Eq.~\eqref{eq:uua_evol}, unlabeled preferential attachment (UPA) trees in Eq.~\eqref{eq:upa0_evol}, and unlabeled preferential attachment with additive shift $\delta$ ($\mathrm{UPA}(\delta)$) in Eq.~\eqref{eq:upa_evol}. By introducing $P_0:=0$ and $R_{-1}:=0$, we may write the equations for $\mathbb{M}$ as
\eq{
\dot{M}_{\ell }&=P_{\ell + 1}-(P_{\ell }+R_{\ell })+R_{\ell - 1}+\delta_{\ell,1}\sum_{\ell'\ge 1}P_{\ell' }, \ \  \ell\ge 0.
}

This description is of interest (i) as an exact descriptions of leaf-statistics in leaf-symmetric growth processes, (ii) as an approximate description of approximately leaf-symmetric growth processes (few nonlocal symmetries), and (iii) as an exact description of {\it approximate models} which themselves account only for local (leaf-) symmetries. Additionally, in preferential attachment settings, we consider approximate models in which degree is replaced by a leaf-degree based proxy (namely, leaf-degree plus one; see Sec.~\ref{sssec:upa0_eqs}). Below, we apply this approach to characterize unlabeled growing tree models.

\subsection{Leaf-degree description of unlabeled tree growth}\label{sec:leaf_symm_method}

To describe how the leaf-statistics $(L,\mathbb{M})$ evolve in each model, we first construct the probabilities $(P_{\ell })_{\ell\ge 1}$ and $(R_{\ell })_{\ell\ge 0}$. This requires summing the attachment kernels associated with all types of attachments to non-leaves and leaves, indexed by $\ell$. Then by applying Eqs.~\eqref{eq:evol_general}, we obtain update equations for the mean values of $(L,\mathbb{M})$, which we assume their true stochastic values are close to. Note, however, that the exact description is by a discrete set of stochastic updates by Eqs~\eqref{eq:update_leafattachment},~\eqref{eq:update_nonleafattachment}, with associated probabilities $((P_{\ell })_{\ell\ge 1},(R_{\ell })_{\ell\ge 0})$ recomputed after each update $n\rightarrow n+1$.

In what follows, we will describe how $R_{\ell }$ and $P_{\ell }$ are obtained from $(L,\mathbb{M})$ in each model. The analysis of the resulting master equations is presented in Sec.~\ref{sec:results}.

\subsubsection{Unlabeled uniform attachment}\label{sssec:uua_eqs}

Attaching to a nonleaf class of leaf-degree $\ell$ in UUA leads to a new leaf and an increment of leaf-degree from $\ell$ to $\ell+1$. This event occurs with probability denoted $R_{\ell }$, equaling 
\eq{
\label{eq:Rl_uua}
R_{\ell }^{\mathrm{UUA}}=\frac{M_{\ell}}{Q_0}, \ \  \ell\ge 0,
}
with normalizer $Q_0$ to be computed after considering the remaining attachment possibilities. If instead we attach to leaf among a group of $\ell\ge 1$ (i.e., those leaves adjacent to a leaf-degree $\ell$ neighbor), the attached-to node joins the nucleus as a rank-$1$ node (not a symmetric duplicate of any previous nucleus node, by assumption of leaf-symmetry). Since each class of leaves is adjacent to one unique rank-$1$ node, the number $\ell$-leaf classes is identical to the number of leaf-degree $\ell$ nonleaves. As such,
\eq{
\label{eq:Pl_uua}
P_{\ell }^{\mathrm{UUA}}=\frac{M_{\ell}}{Q_0}, \ \   \ell\ge 1,
}
in stark contrast to what it would be for {\it labeled} growth statistics, namely, $P_{\ell }\propto \ell M_{\ell}$, with factor $\ell$ representing the $\ell$ distinct leaves attachable to. In leaf-symmetric unlabeled growth, said attachment possibilities collapse to just one, leading to probability weighting $P_{\ell }\propto M_{\ell}$. 

The normalizer $Q_0$ in Eqs.~\eqref{eq:Rl_uua},~\eqref{eq:Pl_uua} resulting in satisfaction of Eq.~\eqref{eq:normalization} for UUA is given by
\eq{
Q_0&=\sum_{\ell\ge 1}M_{\ell}+\sum_{\ell\ge 0}M_{\ell}\\
&=2(n-L)-M_0.
}
Eqs.~\eqref{eq:evol_general} then become
\eq{
\label{eq:uua_evol}
Q_0\dot{L}
&=n-L,\\
Q_0\dot{M_0}&= M_1-M_0 ,\\ 
Q_0\dot{M_1}&= M_2-2M_1+n-L ,\\
Q_0\dot{M_{\ell}}&=M_{\ell + 1}-2M_{\ell}+M_{\ell - 1}, \ \  \ell\ge 2.
}
Our analysis of these equations is presented in Sec.~\ref{ssec:uua_results}.

\subsubsection{Unlabeled preferential attachment}\label{sssec:upa0_eqs}

\begin{figure}[t]
    \centering
\includegraphics[scale=0.8]{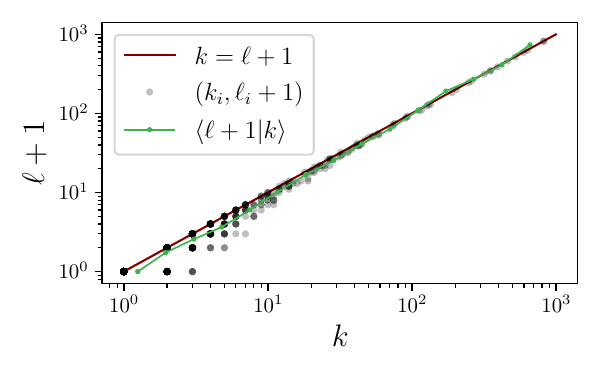}
\\
\includegraphics[scale=0.8]{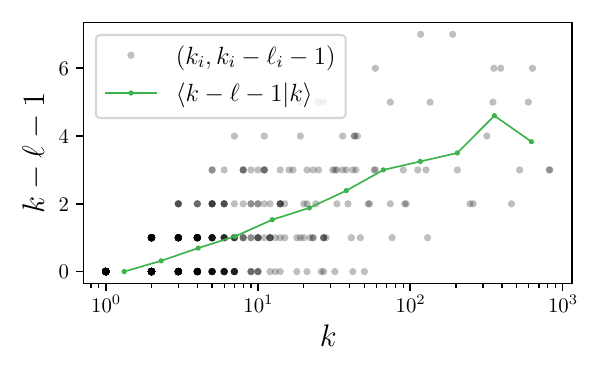}
    \caption{Degree vs leaf-degree plus one in UPA. The mean deviation across all nodes is $\langle k_i-\ell_i-1\rangle\approx 0.056$. The addition $+1$ in $k_i\approx \ell_i+1$ improves the fit by $1$ for every datapoint, since all nodes have at least one nonleaf neighbor. Scattered points represent data from $10$ UPA graphs of size $n=1000$; green dotted curves represent log-binned averages. Upper panel: direct comparison between $\ell_i+1$ and $k_i$, showing that the former closely approximates the latter on the scale of the data values. Lower panel: deviations $k_i-\ell_i-1$, which grow slowly with $k$. }
    \label{fig:deg_leafdeg}
\end{figure}

Models of degree-based preferential attachment have been widely studied in network science. Descriptions of how degree distributions evolve are often obtained by reasoning about the numbers $(N_k)_{k\ge 0}$ of nodes with degree exactly $k$. In contrast, we herein consider leaf-based statistics $(M_{\ell})_{\ell\ge 0}$ as introduced in Sec.~\ref{sec:leaf_based}. To approximate degree-based preferential attachment in a leaf-based description, we require an approximation of {\it degree} in terms of {\it leaf-}degree. Hubs are primarily connected to leaves; $\ell_i/k_i\approx 1$ in the degree and leaf-degree tail. See Fig.~\ref{fig:deg_leafdeg} which displays degree vs leaf-degree in UPA, showing alignment up to sublinearly growing deviations. Rather than $k_i\approx \ell_i$, we approximate degree as
\eq{
k_i\approx \ell_i+1,
}
with the $+1$ accounting for the fact that all vertices have at least one nonleaf neighbor. This choice also correctly reproduces $k_j=1$ for any leaf $j$, since $\ell_j=0$. See the right panels of Fig.~\ref{fig:deg_leafdeg} demonstrating the improvement of $\ell_i+1$ over $\ell_i$ as an approximation of $k_i$. Furthermore, simulations indicate that identical anomalous tail-behavior (powerlaw exponent $\gamma_0\approx 1.84$) is exhibited in the leaf-degree and degree distribution of UPA and as well as its leaf-degree based proxy model. (In contrast, in UUA, $r_{\ell }$ and $p_k$ have different tail-behavior; see Fig~\ref{fig:uua_leafdeg}.)

Under the leaf-degree proxy, the leaf-statistic variables in UPA are described by transition probabilities
\eq{
P_{\ell }^\mathrm{UPA}&=\frac{M_{\ell}}{Q_1}, \ \ell\ge 1;\\ R_{\ell }^\mathrm{UPA}&=\frac{(1+\ell)M_{\ell}}{Q_1}, \ \ell \ge 0; \\
}
with normalization
\eq{
Q_1&=\sum_{\ell\ge 1}M_{\ell}+\sum_{\ell\ge 0}(1+\ell)M_{\ell}\\
&=2n-L-M_0.
}
Note the relation
\eq{
Q_1&=Q_0+L.\\
}
As such, for UPA, Eqs.~\eqref{eq:evol_general} become
\eq{
\label{eq:upa0_evol}
Q_1\dot{L}&= n ,\\
Q_1\dot{M_0}&= M_1 - M_0 ,\\ 
Q_1\dot{M_1}&= M_2-3M_1+n-L , \\
Q_1\dot{M_{\ell}}&= M_{\ell + 1}-(2+\ell)M_{\ell}+\ell M_{\ell - 1} , \ \  \ell\ge 2.
}
Our analysis of these equations is presented in Sec.~\ref{ssec:upa0_results}.

\subsubsection{Unlabeled shifted linear preferential attachment}\label{sssec:upa_eqs}

We define a leaf-degree based proxy model of unlabeled shifted linear preferential attachment, in which $\ell_i+1$ instead of $k_i$ govern attachments. Leaf-attachment and nonleaf-attachment probabilities are of the form
\eq{
P_{\ell }^\mathrm{UPA(\delta)}&=\frac{(1+\delta)M_{\ell}}{Q_\delta}, \ \ell\ge 1;\\ R_{\ell }^\mathrm{UPA(\delta)}&=\frac{(1+\ell+\delta)M_{\ell}}{Q_\delta}, \ \ell \ge 0; \\ 
}
with normalization
\eq{
Q_\delta&=(1+\delta)\sum_{\ell\ge 1}M_{\ell}+\sum_{\ell\ge 0}(1+\ell+\delta)M_{\ell}\\
&=(1+\delta)(2(n-L)-M_0)+L.
}
This leads Eqs.~\eqref{eq:evol_general} to become recursions of the form
\eq{
\label{eq:upa_evol}
Q_\delta\dot{L}&=(1+\delta)(n-L)+L\\
Q_\delta\dot{M}_0&=(1+\delta)(M_1-M_0)\\
Q_\delta\dot{M_1}&= (1+\delta)(n-L+M_2-2M_1)-M_1\\
Q_\delta\dot{M_{\ell}}&=(1+\delta)M_{\ell + 1}-(2+2\delta+\ell)M_{\ell}+(\ell+\delta)M_{\ell - 1}
}
amenable to analytical treatment akin to that for UPA, presented in Sec.~\ref{ssec:upa_results}.

Note the relations
\eq{
Q_\delta&=(1+\delta)Q_0+L&\\
&=Q_1+\delta Q_0.
}

Also note that the $\mathrm{UPA}(\delta)$ model recovers the UUA model in the limit $\delta\rightarrow\infty$, and recovers the UPA model in the limit $\delta\rightarrow 0$. The probability of star graph $S_n$ approaches $1$ as $\delta\downarrow -1$.

\section{Properties of unlabeled growth}\label{sec:results}

Having formulated unlabeled growth and constructed unlabeled analogues to labeled growth models in Sec.~\ref{sec:formulation}, and having developed the leaf-based analytical formalism in Sec.~\ref{sec:leaf_based}, we herein proceed in presenting analytical characterizations of statistical properties of unlabeled growth models. Across models, results reflect a common theme: enhanced degree heterogeneity (and leaf-degree heterogeneity) in unlabeled tree growth in comparison with associated labeled models.

\subsection{UUA analysis}\label{ssec:uua_results}

In UUA (unlabeled uniform attachment, the unlabeled RRT), leaves are observed at higher frequency than in the labeled UA---a fraction $c\approx 0.57$ (obtained below) as opposed to $\frac{1}{2}$. Labeled UA shows higher probabilities of nodes of degrees $2$, $3$, and $4$, whereas degree values of $k>4$ are more probable in UUA. The tail behavior in UUA in fact appears exponential, of the form $p_k\sim c^k$; this remains conjecture, as our approach herein only provides the {\it leaf-degree} distribution $r_\ell$. The methods of Sec.~\ref{sec:leaf_symm_method} allow us to show that $r_\ell\sim(1-c)^\ell$. In contrast, labeled RRTs have degree distribution is $p_k=2^{-k}$, geometric with base $\frac{1}{2}$; furthermore, the labeled RRT has leaf-degree distribution which decays {\it factorially} as $r_\ell\sim 1/(\ell+1)!$; in UUA, both $p_k$ and $r_\ell$ are substantially more heterogeneous.

The heightened heterogeneity in UUA relative to RRTs is also reflected in the star-probability
\eq{
p_n(S_n)=2^{3-n},
}
since at each step $S_{n-1}\rightarrow S_{n}$ for $n\ge 4$, there are two orbits (i.e., ``hub'' and ``leaf'' nodes) and thus a star-retainment probability of $\frac{1}{2}$. This is a much slower decay than $p_n^L(S_n)=2/(n-1)!$ for labeled graphs.

We analyze the growth of UUA via the leaf-based formalism introduced in Sec.~\ref{sec:leaf_based}. We apply an ansatz of linear growth, 
\eq{
L &\approx cn\\
M_{\ell} &\approx m_{\ell }n, \ \   \ell\ge 0,
}
where 
\eq{
c+\sum_{\ell\ge 0}m_{\ell }=1.
}
Then $\dot{L}\approx c$, $\dot{M}_{\ell }\approx m_{\ell }$, and Eqs.~\eqref{eq:uua_evol} become
\eq{
\label{eq:uua_evol_ansatz}
c&=\frac{1-c}{2(1-c)-m_0},\\
m_0&=\frac{m_1-m_0}{2(1-c)-m_0},\\
m_1&=\frac{m_2-2m_1+1-c}{2(1-c)-m_0},
}
and
\eq{
\label{eq:ml}
m_{\ell }&=\frac{m_{\ell + 1}-2m_{\ell }+m_{\ell - 1}}{2(1-c)-m_0}, \ \  \ell\ge 2.
}

\begin{figure}[t]
    \centering
    \includegraphics[width=\linewidth, trim=10 10 0 0]{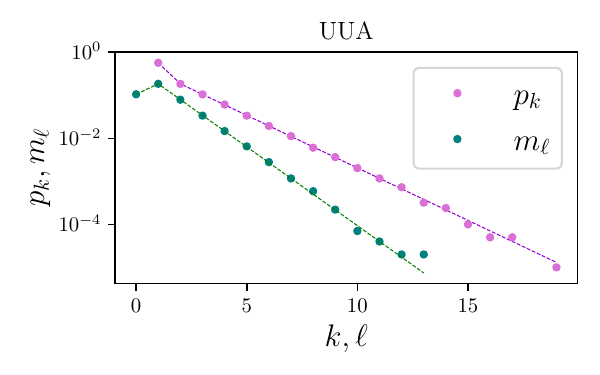}\\ 
    \includegraphics[width=\linewidth, trim=10 10 0 0]{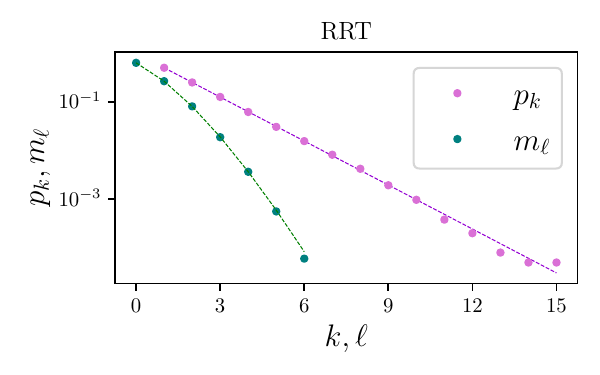}\\ 
    \caption{Degree distribution $p_k$ and leaf-degree fractions $m_{\ell}$ in the leaf-symmetric approximation of UUA, and, for reference, in labeled RRTs. The dashed lines represent analytical predictions. For UUA, $p_k\sim c^k$ (conjecturally) and $m_\ell\sim(1-c)^\ell$ (see Sec.~\ref{ssec:uua_results}); for the RRT, $p_k= 2^{-k}$ and $m_\ell\sim 1/(\ell+1)!$.}
    \label{fig:uua_leafdeg}
\end{figure}

A linear second order recursion with constant coefficients implies a geometric tail, motivating ansatz
\eq{
m_{\ell } = m_1\mu^{\ell-1}, \ \ell\ge 1,
}
from which the third of Eqs.~\eqref{eq:uua_evol_ansatz} becomes
\eq{
\label{eq:m1_mu}
m_1=\frac{(\mu - 2)m_1+1-c}{2(1-c)-m_0},
}
where $(c,m_0,m_1,\mu)$ also satisfy normalization condition
\eq{
\label{eq:norm_urrt}
m_0+\frac{m_1}{1-\mu}=1-c. 
}
To solve the system of equations, we successively write quantities as functions of $c$. For $m_0$ and $m_1$ we obtain
\eq{
\label{eq:m0_m1}
m_0&=(1-c)\left(2-\frac{1}{c}\right)\\
m_1&=\frac{m_0}{c}=\left(\frac{1}{c}-1\right)\left(2-\frac{1}{c}\right).\\ 
}
Solving for $\mu$ in Eq.~\eqref{eq:m1_mu} allows us to write
\eq{
\label{eq:mu_before_cancellation}
\mu&=2(1-c)-m_0+2-\frac{1-c}{m_1}\\
&=1-c+\frac{c^3-c^2+2c-1}{(2c-1)c},
}
where the choice of algebraic form in Eq.~\eqref{eq:mu_before_cancellation} is motivated by what follows. We re-express the normalization condition (Eq.~\eqref{eq:norm_urrt}) in terms of $c$ alone by application of Eqs.~\eqref{eq:m0_m1},~\eqref{eq:mu_before_cancellation} to obtain the cubic
\eq{
\label{eq:uua_cubic}
c^3-c^2+2c-1=0.
}
As such, $c$ is the real root of Eq.~\eqref{eq:uua_cubic}. In particular,
\eq{
\label{eq:c}
c&=\frac{1}{3}\left(1 - 5 \sqrt[3]{\frac{2}{11 + 3 \sqrt{69}}} + \sqrt[3]{\frac{11 + 3 \sqrt{69}}{2}}\right)\\
&=0.56984029...
}
Furthermore, noting that the numerator of the last expression for $\mu$ in Eq.~\eqref{eq:mu_before_cancellation} is identically the cubic of Eq.~\eqref{eq:uua_cubic}, the quantity $\mu$ collapses to the simple form
\eq{
\mu=1-c.
}
Finally, we confirm that indeed this value of $\mu$ is compatible with Eqs.~\eqref{eq:uua_evol_ansatz} for $\ell\ge 2$. Plugging in ansatz $m_{\ell }=m_1\mu^{\ell-1}$ yields
\eq{
\mu = \frac{\mu^2-2\mu+1}{2(1-c)-m_0},
}
which, upon application of $\mu=1-c$ and $m_0=(1-c)(2-1/c)$, can be algebraically rearranged to again obtain the cubic of Eq.~\eqref{eq:uua_cubic}. As such, the complete solution in terms of $c$ is
\eq{
m_0&=(2-1/c)(1-c)\\
m_{\ell }&=(1/c)(2-1/c)(1-c)^{\ell}, \  \ell\ge 1,
}
or more compactly, for all $l\ge 0$,
\eq{
m_{\ell }&=\frac{1}{c}\left(2-\frac{1}{c}\right)\left(\frac{c}{1-c}\right)^{\delta_{\ell,0}}(1-c)^{\ell}, \  \ell\ge 0.
}
This matches numerical simulations; see Fig.~\ref{fig:uua_leafdeg}. The absolute numbers at size $n$ are around $L\approx cn$ and $M_{\ell}\approx m_{\ell }n$. Numerically, the original variables $(c,m_0,m_1,\mu)$ are approximately
\eq{
c &= 0.56984...,\\
m_0 &= 0.10544...,\\
m_1 &= 0.18503...,\\
\mu &= 0.43015....
}

This leaf-fraction computation also provides $p_1=c$, with $p_k$ denoting the degree distribution. In the leaf-symmetric proxy model, the remainder of the degree distribution appears to be exactly geometric, but with some base exceeding the RRT value of $\frac{1}{2}$. We conjecture $p_k\sim c^k$. By obtaining $m_0$ we have also found the {\it protected fraction}, i.e., the fraction of nodes with rank exceeding $1$, where rank is the distance to the closest leaf. We may further compute the fraction of {\it rank-one} nodes, i.e., the number of neighbors of leaves, as
\eq{
\sum_{\ell\ge 1}m_{\ell }&=(1/c)(2-1/c)\sum_{\ell\ge 1}(1-c)^{\ell}\\
&=\frac{(1-c)(2c-1)}{c^3}=0.32471....
}
These provide just a few examples of the analytically tractable nature of leaf-symmetric tree growth, in the unlabeled analog (UUA) of the simplest labeled tree growth model (RRT); many avenues of inquiry remain open, such as determination of the degree distribution, conjectured to have the form
\eq{
p_k=\delta_{k,1}c+(1-\delta_{k,1})(1/c-1)^2c^{k}.
}

\subsection{UPA analysis}\label{ssec:upa0_results}

In the UPA model, linear degree-based preferential attachment combines with the symmetry effects of unlabeled growth to produce extraordinary degree heterogeneity. In particular, the leaf-fraction $c=L/n$ approaches $1$, leaving a sublinearly growing ``nucleus'' consisting of $\sim n^\xi$ nonleaves \cite{gabel2014highly, krapivsky2017emergent}, where $\xi \in(0,1)$. Within the nucleus, a powerlaw-tailed degree distribution is exhibited: $p_k\sim k^{-(1+\xi)}$, with anomalously subquadratic tail-exponent $\gamma_0=1+\xi<2$, and cutoff at $n$. We obtain the value of $\xi\approx 0.84$ by analyzing the recursion arising in the leaf-based formalism (see Appendix~\ref{app:exponent_finding}). See Fig.~\ref{fig:UPA} for visualization of a random UPA graph and simulation data of the degree and leaf-degree distributions in the leaf-symmetric, leaf-based proxy model.

Similar anomalously heterogeneous random trees were also reported in the isotropic redirection model \cite{krapivsky2017emergent}; it has also been observed in degree-based redirection \cite{gabel2014highly}. Related but distinct phenomena include condensation effects in superlinear preferential attachment \cite{krapivsky2001organization} and linear preferential attachment with heterogeneous fitnesses \cite{bianconi2001bose}. Other studies have considered models with powerlaw-degree tails of exponent less than two \cite{seyed2006scale, timar2016scale, d2007power} but not restricted to trees, instead permitting a diverging average degree. See Fig.~\ref{fig:UPA} for visualization of a random UPA graph.

\begin{figure}[t]
    \centering
    \includegraphics[width=0.9\linewidth,trim=15 15 15 15,clip]{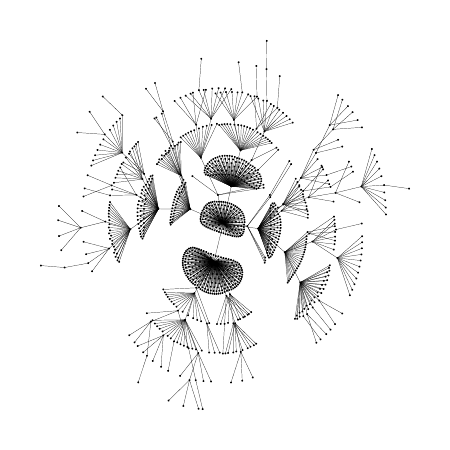}\\
    \includegraphics[width=\linewidth,trim=10 0 0 0,clip]{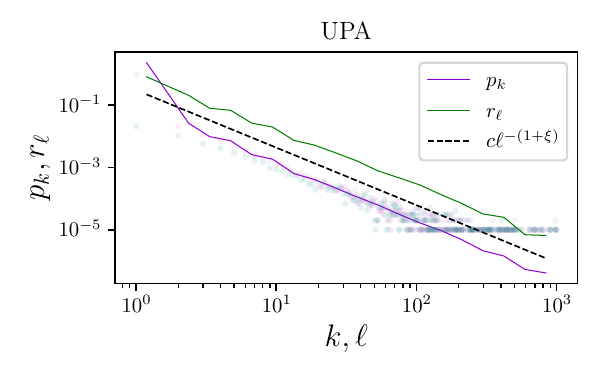}
    \caption{Random UPA tree at size $n=500$. Lower panel: log-binned degree distribution from $100$ realizations at $n=1000$ of UPA in the leaf-symetric approximation; unbinned frequencies are displayed semi-transparently; pure powerlaw decay with exponent $\gamma_0\approx 1.84$ obtained as $\gamma_0=1+\xi$ where $\xi$ solves Eq.~\ref{eq:xi_equation}.}
    \label{fig:UPA}
\end{figure}

The exotic properties of UPA are further demonstrated by the scaling of specific graph probabilities. Recall, in labeled PA, the star-probability decays as $p_n^L(S_n)\sim 2^{-n}$. In UPA, the star-probability is
\eq{
p_n(S_n)=\prod_{j=2}^{n-2}\frac{j}{1+j}=\frac{1}{n-1}.
}
Such slowly decaying star-probabilities were observed in the isotropic redirection model \cite{krapivsky2017emergent}, degree-based redirection \cite{gabel2014highly}, and in (labeled) nonlinear preferential attachment in the superlinear regime at $\alpha=2$ \cite{krapivsky2001organization}. The UPA model is further addressed analytically in Sec.~\ref{sec:leaf_symm_method}.

We analyze Eqs.~\eqref{eq:upa0_evol} to characterize the UPA model in terms of leaf-statistic variables. Numerical evidence of the leaf-proliferation phenomenon ($L/n\rightarrow 1$) motivates ansatz of the form
\eq{
\label{eq:upa_ansatz}
L&\approx n-bn^\xi,\\
M_{\ell}&\approx w_{\ell }n^\xi, \ \  \ell\ge 0,
}
with $\xi\in(0,1)$, $b>0$, $w_{\ell }>0$,  and $\sum_{\ell\ge 0}w_{\ell }=b$. Derivatives with respect to $n$ can be expressed as $\dot{L}=1-\xi b n^{\xi-1}$ and $\dot{M_{\ell}}=\xi w_{\ell } n^{\xi-1}$. Under the ansatz of Eq.~\eqref{eq:upa_ansatz}, the normalizer becomes
\eq{
Q & = 2n-L-M_0 \\
& = n+(b-m_0)n^\xi.
}
For $n\gg 1$ we Taylor expand with $n^{\xi-1}$ as a small parameter. The evolution equations then reduce to the form
\eq{
\label{eq:upa0_evol_leafdeg}
\xi  &= 1-r_0 ,\\
\xi r_0 &= r_1-r_0,\\ 
\xi r_1&= 1+r_2-3r_1\\
\xi r_{\ell }&= r_{\ell + 1}-(2+\ell )r_{\ell }+\ell r_{\ell - 1}, \ \  \ell\ge 2.
}
where we have eliminated $b$ by switching to the normalized leaf-degree distribution $r_{\ell }:=w_{\ell }/b$.

A first observation is the relation
\eq{
\xi =1-r_0 ,
}
with $r_0:=w_0/b$ the {\it protected fraction} of nonleaves. This in turn defines the powerlaw tail-exponent relation $\gamma_0 =2-r_0$; a powerlaw-tailed solution is verified by application of ansatz $r_{\ell }\sim \ell^{-\gamma_0 }$ for some $\gamma_0\in(1,2)$ to the fourth equation of Eqs.~\eqref{eq:upa0_evol_leafdeg}. For $\ell\gg 1$, the recursion then behaves as
\eq{
\left(1-\frac{1}{\ell}\right)^{-\gamma_0}\simeq 2+\xi+\ell -\ell\left(1-\frac{1}{\ell}\right)^{-\gamma_0};
}
applying $(1+x)^{-\gamma_0}\approx 1-\gamma_0 x$ for $|x|\ll 1$, and dropping terms of order $1/\ell$, we find the required exponent:
\eq{
\gamma_0=1+\xi.
}
In Appendix~\ref{app:exponent_finding} we show that the only value of $\xi$ for which Eqs.~\eqref{eq:upa0_evol_leafdeg} admit an asymptotically decaying solution must solve the following continued fraction equation:
\eq{
\label{eq:xi_equation}
3+\xi-\frac{1}{1-\xi^2}=\frac{2}{\xi+4+\cfrac{3}{\xi+5+\cfrac{4}{\xi+6+\cdots}}},
}
the numerical value of which is $0.84355...$; the leaf-degree distribution behaves as
\eq{
r_{\ell }\sim \ell^{-\gamma_0},
}
with $\gamma_0=1+\xi\approx 1.84$. The degree distribution $p_k$ follows similar tail behavior: $p_k\sim k^{-\gamma_0}$, as expected based on the approximate relationship $k_i\approx\ell_i+1$ (see Fig.~\ref{fig:deg_leafdeg}). The anomalous exponent $\gamma_0<2$ and vanishing nonleaf-fraction is reminiscent of the isotropic redirection model \cite{krapivsky2017emergent}.

\subsection{$\mathrm{UPA}(\delta)$ analysis}\label{ssec:upa_results}

\begin{figure*}[t]
    \centering
    \includegraphics[width=0.7\linewidth]{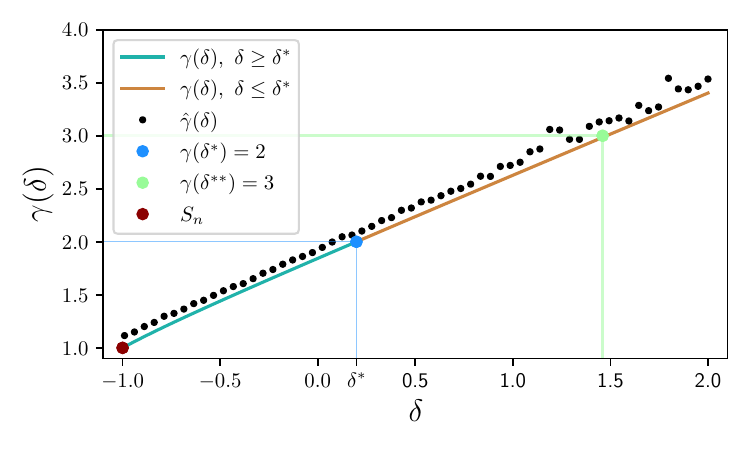}
    \caption{Powerlaw tail exponent of the leaf-degree distribution in stochastically simulated in the leaf-symmetric and leaf-degree based proxy model for $\mathrm{UPA}(\delta)$. Data is from $240$ trials at size $n=2\times 10^4$, with finite size effects yielding slight overestimation across $\delta$. The curve $\gamma(\delta)$ is obtained in the anomalous regime as $\gamma(\delta)=1+\xi(\delta)$ by solution of Eq.~\eqref{eq:xi_implicit} for $\xi(\delta)$, and in the linear scaling regimes by Eq.~\eqref{eq:gamma_of_c} for $\gamma(\delta)$ in terms of $c(\delta)$, with $c(\delta)$ obtained as the solution of Eq.~\eqref{eq:c_of_delta_implicit}. The curve is well-approximated by the tangent at $\delta=0$ of the form $\gamma=1+(1+\delta)(1-r_0(0))$, with $r_0(0)$ the protected fraction in $\mathrm{UPA}$; this is akin to the linear relationship $\gamma_L=3+\delta_L$ arising in the labeled model $\mathrm{PA}(\delta_L)$. The key difference is that the unlabeled model can parametrically access the anomalous regime $\gamma\in(1,2]$, despite trees having finite first moment of degree.}
\label{fig:gamma_of_delta}
\end{figure*}

In $\mathrm{UPA}(\delta)$, we find a powerlaw-tailed degree distribution $p(k)\sim k^{-\gamma}$ with exponent $\gamma(\delta)\ge 1$ being a linear function of $\delta$. As such, there exists a $\delta=\delta^\star>0$ such that scale-free leaf-proliferation ($\gamma\le 2$) is exhibited for $\delta\le \delta^\star$, and non-anomalous scaling ($\gamma>2$) is exhibited for $\delta>\delta^\star$, with a crossover point with $\gamma(\delta^\star)=2$. Numerical simulations indicate this behavior (see Fig.~\ref{fig:gamma_of_delta}). The analytical techniques of Sec.~\ref{sec:leaf_symm_method} also reproduce this behavior, as we show below.

As an illustration of $\mathrm{UPA}(\delta)$ behavior, consider the star probability is
\eq{
p_n(S_n)&=\prod_{j=2}^{n-2}\frac{j+\delta}{j+1+2\delta}\\
&\simeq \frac{\Gamma(3+2\delta)}{\Gamma(2+\delta)}n^{-(1+\delta)}.
}
Additionally, we obtain the probability $P_{m,n}$ of a pair of connected nodes attached to $m$ leaves and $n$ leaves, respectively, by analyzing the recursion
\eq{
P_{m,n}&=\frac{(m+\delta)P_{m-1,n}+(n+\delta)P_{m,n-1}}{m+n+1+4\delta}.\\
}
Supposing $m,n\gg 1$, we take a continuum approximation and Taylor expand to obtain a solution of the form
\eq{
P_{m,n}=\frac{A}{(mn)^{\frac{1}{2}+\delta}},
}
for some $A>0$. In the case of UPA ($\delta=0$) we have $P_{m,n}\sim 1/\sqrt{mn}$, the same scaling as observed in the isotropic redirection model \cite{krapivsky2017emergent}.

The smallest non-leaf symmetry is in a line graph of size five (see Fig.~\ref{fig:symmetry_schematic}(i)). The probability of that is, in $\mathrm{UPA}(\delta)$, the probability of two leaf-attachments in a row after initialization at $n=3$, which is $\frac{(1+\delta)^2}{8}$. The probability of this early nonlocal symmetry breaking becomes negligible as $\delta\downarrow -1$, reflecting localized nature of symmetries in heterogeneous trees.

Turning to the leaf-based formalism, we describe how the additive shift $\delta$ alters the model behavior in comparison with UPA (Sec.~\ref{ssec:upa0_results}). We also compare the behavior to {\it labeled} shifted linear preferential attachment trees \cite{dorogovtsev2000structure}, also parameterized by $(n,\delta)$. The key result is a modification from the labeled model's powerlaw tail exponent $\gamma_{L}(\delta)=3+\delta$ to one of the form $\gamma(\delta)=\gamma_0+\xi\delta$, where $\xi=\gamma_0-1\approx 0.84$ is the UPA scaling exponent (see Sec.~\ref{ssec:upa0_results}). The labeled and unlabeled models respectively approach $\gamma\downarrow 2$ and $\gamma\downarrow 1$ as $\delta\downarrow -1$, with the latter scenario accompanied by leaf-proliferation for all $\gamma\le 2$. In what follows, we analyze the set of equations Eqs.~\eqref{eq:upa_evol} derived for leaf-based variables under the leaf-degree based approximation of degree $k_u\approx \ell_u+1$. We first focus on the leaf-proliferating regime $\delta\le \delta^*$ ($\gamma(\delta)\le 2$), where the nonextensive scaling ansatz is applicable; we then consider the linear scaling regime $\gamma(\delta)\ge 2$, via a linearly scaling ansatz. The two approaches align at the critical point $\delta=\delta^*$.

Under an analogous scaling ansatz to that of Sec.~\ref{sssec:upa0} but with $\xi=\xi(\delta)$, we introduce $b$ and $\{w_{\ell }\}_{\ell\ge 0}$ for which $L\approx n-bn^{\xi(\delta)}$ and $M_{\ell}\approx w_{\ell }n^{\xi(\delta)}$, with $\sum_{\ell\ge 0}w_{\ell }=b$. This ansatz is only consistent in the regime $\xi(\delta)\le 1$, with linear growth recovered otherwise. For convenience, we denote the nucleus size $F:=n-L=\sum_{\ell\ge 0}M_{\ell}$ and note that $\dot{F}=1-\dot{L}$.

Approximating to first order in $n^{\xi-1}$, and eliminating $b$ by $r_{\ell }:=w_{\ell }/b$, Eqs.~\eqref{eq:upa_evol} become
\eq{
\label{eq:upa_evol_approx}
\xi & =(1+\delta)(1-r_0),\\
\xi r_0 &= (1+\delta)(r_1-r_0),\\
\xi r_1 &= (1+\delta)(1+r_2-2r_1)-r_1,\\
\xi r_{\ell } &= (1+\delta)r_{\ell + 1}-(2+2\delta+\ell )r_{\ell }+(\ell+\delta)r_{\ell - 1}, \  \ell\ge 2,
}
with the special case of $\delta=0$ recovering Eqs.~\eqref{eq:upa0_evol_leafdeg}.

We first verify that application of ansatz $r_{\ell }\simeq C\ell^{-\gamma(\delta)}$ again provides a valid solution, and that the relationship $\gamma(\delta)=1+\xi(\delta)$ is maintained. From the 2nd order recursion (the fourth of Eqs.~\eqref{eq:upa_evol_approx}), we have
\eq{
2(1+\delta)+\xi(\delta)+\ell -(\ell+\delta)&\left(1-\frac{1}{l}\right)^{-\gamma(\delta)}\\&\approx (1+\delta)\left(1+\frac{1}{l}\right)^{-\gamma(\delta)},
}
or, for $\ell\gg 1$,
\eq{
\gamma(\delta)&\simeq 2(1+\delta)+\xi(\delta)-\delta- (1+\delta)\\
&=1+\xi(\delta).\\
}
In Appendix~\ref{app:exponent_finding} we show that $\xi(\delta)$ solves the following equation:
\begin{widetext}
\eq{
\label{eq:xi_implicit}
\frac{3+2\delta+\xi(\delta)}{1+\delta}-\frac{(1+\delta)^2}{(1+\delta)^2-\xi(\delta)^2} &
=\frac{2+\delta}{\xi(\delta)+4+2\delta+\cfrac{3+\delta}{\xi(\delta)+5+2\delta+\cfrac{4+\delta}{\xi(\delta)+6+2\delta+\cdots}}}.
}
\end{widetext}
Using the first of Eqs.~\eqref{eq:upa_evol_approx}, we may also write
\eq{
\label{eq:gam_of_delta}
\gamma(\delta) &= 1+(1+\delta)(1-r_0),\\
&=(2-r_0)+(1-r_0)\delta,
}
with $r_0$ denoting the protected fraction of nonleaves ($r_0:=w_0/b=M_0/F$). Recall, the analogous relationship in {\it labeled} $\mathrm{PA}(\delta)$ is $\gamma_L=3+\delta_L$ \cite{van2014random}. See Fig.~\ref{fig:gamma_of_delta} for numerical evidence of Eq.~\eqref{eq:gam_of_delta} as measured from direct simulations of $(L,(M_{\ell})_{\ell\ge 0})$ with the leaf-degree proxy for $\mathrm{UPA}(\delta)$. Eq.~\eqref{eq:gam_of_delta} can be well approximated by the linear dependency $\gamma(\delta)\approx 1+(1+\delta)r_0(0)$, with $r_0(0)$ denoting the protected fraction for $\mathrm{UPA}$ (see Fig.~\ref{fig:gamma_of_delta}).

There is a positive threshold value $\delta^*>0$ for which $\gamma(\delta^*)=2$. By setting $\xi=1$ in Eq.~\eqref{eq:xi_implicit} we find that $\delta^*$ solves the following equation:
\eq{
\frac{4+2\delta}{1+\delta}-\frac{(1+\delta)^2}{(1+\delta)^2-1}&=\frac{\delta+2}{5+2\delta+\cfrac{\delta+3}{6+2\delta+\cfrac{\delta+4}{7+2\delta+\cdots}}},
}
from which we obtain $\delta^*=0.19831...$; for $\delta>\delta^*$, numerical evidence suggests that linear scaling holds. This motivates application of an ansatz akin to that for $\mathrm{UUA}$: $L\simeq cn,M_\ell \simeq m_\ell n$, $\ell\ge 0$, with $\sum_{\ell}m_\ell=1-c$. In this case, Eqs.~\eqref{eq:upa_evol} become
\eq{
\label{eq:upa_linear}
q_\delta c &= (1+\delta)(1-c)+c,\\
q_\delta m_0 &= (1+\delta)(m_1-m_0),\\
q_\delta m_1 &= (1+\delta)(1-c+m_2-2m_1)-m_1,\\
q_\delta m_\ell &= (1+\delta)m_{\ell+1}-(2+2\delta+\ell)m_\ell+(\ell+\delta)m_{\ell-1}, \\
}
with the last equation above holding for $\ell\ge 2$, and where $q_\delta=Q_\delta/n$ is given by
\eq{
q_\delta=(1+\delta)(2(1-c)-m_0)+c.
}
From the first of Eqs.~\eqref{eq:upa_linear}, we obtain the relationship between $c$ and $m_0$, leading to
\eq{
m_0&=(1-c)\left(\frac{2c-1 }{c}-\frac{1}{1+\delta}\right).
}
This yields $q_\delta$ as a function of $c$:
\eq{
\label{eq:qdelta}
q_\delta= 1+\frac{(1+\delta)(1-c)}{c}.
}
The second equation yields an expression for $m_1$:
\eq{
m_1&=\frac{(q_\delta+1+\delta)m_0}{1+\delta}\\
&=\left(\frac{1}{1+\delta}+\frac{1}{c}\right)(1-c)\left(\frac{2c-1 }{c}-\frac{1}{1+\delta}\right).
}

The next equation can be used to derive $m_2$, and so on; the value of $c(\delta)$ is determined by the asymptotic boundary condition $m_\infty=0$; in Appendix~\ref{app:exponent_finding} we show that $c(\delta)$ satisfies
\begin{widetext}
\eq{
\label{eq:c_of_delta_implicit}
\left(1+\frac{2}{1+\delta}+\frac{1}{c(\delta)}\right)-\left(\frac{1}{1+\delta}+\frac{1}{c(\delta)}\right)^{-1}\left(2-\frac{1}{1+\delta}-\frac{1}{c(\delta)}\right)^{-1}=\frac{2+\delta  }{4+2\delta+q_\delta +\cfrac{3+\delta}{5+2\delta+q_\delta +\cfrac{4+\delta }{6+2\delta+q_\delta+\cdots}}},
}
\end{widetext}
with $q_\delta$ given by Eq.~\eqref{eq:qdelta}. The leaf-degree distribution has a powerlaw tail with exponent $\gamma>2$ in the regime $\delta>\delta^*$; applying ansatz $m_\ell=C\ell^{-\gamma}$ to the last of Eqs.~\eqref{eq:upa_linear},
\eq{
q_\delta &=(1+\delta)(1+1/\ell)^{-\gamma}\\
&-(2+2\delta+\ell)+(\ell+\delta)(1-1/\ell)^{-\gamma};\\
}
in the $\ell\gg 1$ regime, dropping terms of order $1/\ell$, we obtain
\eq{
\gamma(\delta)=1+q_\delta.
}
Application of Eq.~\eqref{eq:qdelta} yields
\eq{
\label{eq:gamma_of_c}
\gamma(\delta)=2+\frac{(1+\delta)(1-c)}{c}.
}
This aligns with the anticipated breakdown in the linear scaling ansatz at $\gamma(\delta^*)=2$: it happens precisely at $c(\delta^*)=1$, i.e., when the fraction of nodes that are nonleaves becomes vanishingly small. In the $\gamma\rightarrow\infty$ limit, the value of $c(\delta)$ approaches that of $\mathrm{UUA}$ (Eq.~\eqref{eq:c}). We may also define $\delta^{**}$ as the value resulting in $\gamma(\delta^{**})=3$, i.e., reproducing the tail-behavior of {\it labeled} PA.

Finally, we briefly discuss the critical model $\mathrm{UPA}(\delta^*)$. It has leaf-degree distribution with powerlaw tail of exponent $\gamma(\delta^*)=2$, which under linear scaling would induce diverging average degree ($\bar{k}\sim \log n$), but since $\bar{k}\rightarrow 2$, leaf-proliferation is induced instead. The nucleus is of size $\mathcal{N}$ and has average degree $\bar{k}_{\mathcal{N}}\sim \log n$. As such, 
\eq{
\bar{k}&=\frac{\mathcal{N}}{n}\bar{k}_{\mathcal{N}}+\left(1-\frac{\mathcal{N}}{n}\right)1\simeq 2,
}
from which
\eq{ \mathcal{N}=\frac{n}{\bar{k}_{\mathcal{N}}-1}\sim\frac{n}{\log n},
}
i.e., the nucleus is of barely vanishing relative size $1/\log n$. This marginal case is what allows for $\xi(\delta^*)=1$ (since $n/\log n$ is linear up to logarithmic factors), while also simultaneously allowing $c(\delta^*)=1$ (since a fraction $1-O(1/\log n)$ of nodes are leaves).

\section{Discussion}\label{sec:discussion}

In this work, we have formulated growth models for unlabeled trees, examining unlabeled versions of widely-studied labeled growth models. Unlabeled growth rules are probability distributions over {\it structurally distinct} outcomes. We constructed unlabeled versions of uniform and preferential attachment, in the models $\mathrm{UUA}$, $\mathrm{UPA}$, and $\mathrm{UPA}(\delta)$. We developed a leaf-based analytical formalism capable of describing unlabeled growth in the leaf-symmetric regime. We explored the basic phenomenology of these models and found that while similarities are present between unlabeled and labeled growth models, substantial differences can also emerge (see Sec.~\ref{sec:results}). These differences manifest primarily through enhanced degree heterogeneity in the unlabeled setting. In particular, in $\mathrm{UUA}$, the unlabeled version of random recursive trees, we found that the degree distribution is geometric with base $c=0.56984...$ in contrast with labeled $\mathrm{RRT}$ value of $\frac{1}{2}$, and that the leaf-degree distribution is geometric with base $1-c$, in contrast with the labeled $\mathrm{RRT}$ leaf-degree distribution which decays factorially. The $\mathrm{UPA}$ model, an unlabeled analogue of the $\mathrm{BA}$ model, has anomalous scaling and leaf-proliferation, with $n-L\sim n^\xi$, and powerlaw tailed degree and leaf-degree distributions of exponent $\gamma_0=1+\xi=1.84355...$ in contrast with the tail exponent $3$ and linear scaling behavior of the $\mathrm{BA}$ model. Finally, in $\mathrm{UPA}(\delta)$, unlabeled preferential attachment with additive shift $\delta$, we found powerlaw degree and leaf-degree tail exponent $\gamma(\delta)\in(1,\infty)$, with anomalous scaling and leaf-proliferation for $\delta\le \delta^*$, and linear scaling for $\delta>\delta^*$, where the critical additive shift value $\delta^*=0.19831...$ is defined by $\gamma(\delta^*)=2$. These results, alongside the establishment of unlabeled growth models in the first place, and the leaf-based formalism developed herein, are the main contributions of this work.

Many questions remain open and numerous extensions are possible. A variety of properties have been analytically characterized in {\it labeled} growing networks, such as the maximal depth \cite{devroye2012depth}, the maximum degree \cite{goh2002limit}, the entropy rate \cite{zhao2011entropy}, and path lengths \cite{dobrow1999total}; all of these are well-posed properties in the unlabeled growth setting. Other properties of interest have been obtained in {\it non-growing} unlabeled random networks \cite{olsson2023distribution}; the analogous properties in unlabeled growth models are open for study. Graph limits have been examined for labeled random graph models \cite{lovasz2006limits, borgs2019L}, including (nongrowing) unlabeled trees \cite{stufler2019continuum, jin2020graph}, and (labeled) growing trees \cite{borgs2011limits, JansonSeverini2013}; what are the limits of growing unlabeled trees? Other unlabeled growth mechanisms could also be formulated, e.g., involving concurrent addition and removal of vertices \cite{bauke2011topological}, copying-based growth \cite{krapivsky2005network} or internal link-formation \cite{callaway2001randomly, dorogovtsev2001anomalous}. 

The further exploration of unlabeled approaches in network science could yield theoretical developments and potential future applications. For example, construction of models in unlabeled sample spaces other than simple trees, such as unlabeled hypergraphs \cite{qian2014enumeration}, obtainment of estimators and inference algorithms for parameter estimation in unlabeled network models \cite{peixoto2013parsimonious}, unlabeled and labeled model comparisons on real-world datasets, among many other topics. These would be facilitated better by development of faster numerical methods for unlabeled sampling algorithms \cite{colbourn1981linear, babai1983canonical, buss1997alogtime, wormald1987generating}. We emphasize that despite the relative unfamiliarity of unlabeled random graphs in network science literature, the standard approaches of modeling and data analysis are equally applicable in the setting of unlabeled graphs. The associated theoretical challenges call for new techniques which may in turn have broader applicability, as is the case for the leaf-symmetric growth formalism that we developed in Sec.~\ref{sec:leaf_based}. We note that in many models of network growth, labels indeed endow information pertinent to attachment, e.g., associated with aging \cite{zhu2003effect}, fitness \cite{bianconi2001competition}, or geometry \cite{zuev2015emergence}. The interplay of the latter mechanisms and unlabeled growth is an interesting open problem. 

The properties of the tree growth models considered herein demonstrate that label-bias can have huge effects in some commonplace modeling settings, namely, undirected tree growth with and without preferential attachment. Unlabeled sample spaces don't factorize reductively into configurations of elementary labeled variables (e.g., pairwise adjacency-indicators), instead encoding the global pattern of connectivity. The nature of unlabeled growth, intrinsically nonlocal, permits consideration of growth mechanisms involving holistic network information. Such mechanisms are plausible for macroscopic networks in which emergent phenomena are the rule rather than the exception. Furthermore, in various scenarios, networks may be viewed as inherently unlabeled. For instance, growing causal sets in quantum gravity~\cite{rideout1999classical, surya2025causal}, where the unlabeled nature of networks is dictated by their Lorentz invariance. Other cases include networks of fractures \cite{https://doi.org/10.1029/JB090iB04p03087,andresen2013topology}, plant physiology \cite{prusinkiewicz2012algorithmic,https://doi.org/10.1111/j.1365-3040.2010.02273.x}, geophysical and biological flow networks \cite{10.1063/1.4908231,PhysRevE.99.012321,PhysRevFluids.7.013101}, road networks \cite{FREIRIA201555,reza2024road}, among others, where nodes serve as junctions for links, rather than as individualized actors with identifying features. Regardless, {\it any} network dataset may be viewed in the unlabeled representation and can be analyzed via unlabeled models; this choice is natural when network {\it structure} is the primary focus. We hope that future research addresses the question: which real-world networks are less surprising in unlabeled network models than in their labeled counterparts? 

\begin{acknowledgments}

We thank S.~Redner and K.~Denton for helpful discussions and feedback. H.H.~acknowledges support from the Santa Fe Institute. D.K.~acknowledges support from NSF Grant No.~CCF-2311160.

\end{acknowledgments}

\bibliographystyle{apsrev4-2}

\appendix

\section{Random unlabeled graphs}\label{app:random_unlabeled}

Unlabeled graphs encode network {\it structure}, whereas labeled graphs encode {\it who is connected to whom}. Unlabeled graphs have been present throughout the history of graph theory \cite{cayley1874lvii, hanlon1985chromatic, naor1990succinct}. This includes the history of {\it random} graphs \cite{schwenk1977asymptotic}, with random unlabeled graphs predating the modern era of network science \cite{luczak1991deal}. Unlabeled random graphs are simply probability distributions over sets of unlabeled graphs. One class of unlabeled random graph is the {\it delabeled} version of a labeled random graph \cite{gaudio2022shotgun, kontoyiannis2021symmetry}, wherein the probability of unlabeled graph $U$ is 
\eq{
P_D(U)=\sum_{G:U_G=U}P_L(G),
}
with $P_L(G)$ denoting a labeled graph model's probability distribution and with $U_G$ denoting the unlabeled structure of a labeled graph $G$. If the model $P_L(G)$ is {\it exchangeable} \cite{lauritzen2018random}, then
\eq{
\label{eq:delabeled_exchangeable}
P_{D}(U)=|\mathrm{Iso}(G_U)|P_L(G_U),
}
with $\mathrm{Iso}(G_U)$ the number of possible labelings of $G_U$, where $G_U$ denotes any labeled version of $U$. 

We note the inverse proportionality $|\mathrm{Aut}(G)|=n!/|\mathrm{Iso}(G)|$ with $\mathrm{Aut}(G)$ the set of automorphisms (network symmetries \cite{garlaschelli2010complex}) of labeled graph $G$. That is, more symmetrical unlabeled graphs are probabilistically suppressed. This may be pertinent to empirical modeling given the symmetries observed in real-world networks \cite{garrido2011symmetry}. 

As an example of labeled, unlabeled, and delabeled models, consider the canonical Erd\H{o}s-R\'enyi model $G(n,p)$ \cite{erdos1959random}. It is the maximum entropy distribution of labeled $n$-node graphs of given expected edge-density $p$ \cite{park2004statistical}; the {\it unlabeled} canonical Erd\H{o}s-R\'enyi model $U(n,p)$ is the maximum-entropy distribution of {\it unlabeled} graphs of edge-density $p$, having an associated Gibbs form \cite{jaynes1957information}. The delabeled Erd\H{o}s-R\'enyi model, denoted $D(n,p)$ \cite{gaudio2022shotgun}, has graph distribution in the form of Eq.~\eqref{eq:delabeled_exchangeable}. In contrast to $G(n,p)$, analytical tractability of $U(n,p)$ is hindered by challenges in unlabeled graph enumeration \cite{wright1973probability, wright1974graphs, promel1987counting}. Nevertheless, recent work has characterized $U(n,p)$ and demonstrated ensemble nonequivalence to $G(n,p)$ \cite{evnin2025ensemble}. Numerous difficulties arise in handling unlabeled random graphs, many of which can be traced to the problem of graph isomorphism \cite{grohe2020graph}.

\section{Labeled growth models}\label{app:labeled_growth}

Herein we review the standard formulation of labeled growing trees. The newly arriving vertex ``$n$'' can attach to any of the pre-existing vertices $\{1,...,n-1\}$. Not all labeled trees may be generated in this way, but rather only those consistent with a valid growth history (a.k.a. {\it increasing} trees \cite{baur2016percolation}). E.g., a $3$-node line graph $L_3$ labeled as $(1,2,3)$ is growable, but isn't growable if labeled as $(1,3,2)$. The number of increasing trees is $(n-1)!$.

For a given labeled tree $G$ of size $n-1$, we let the set of graphs growable from it be denoted $\mathcal{G}(G)$. A {\it growth rule} is any distribution  $P_n(G'|G)$ assigning probability to each $G'\in\mathcal{G}(G)$. By being labeled, $|\mathcal{G}(G)|=n-1$, and we may index the possible growth outcomes according to which labeled node $i\in V(G)$ is attached to.

\subsection{Random recursive trees}\label{sssec:rrt}

Random recursive trees (RRTs) are the simplest labeled growth rule: a uniform distribution over the set $\mathcal{G}(G)$, i.e., a constant attachment kernel $f(G,G')=1$. Thus in growth step $n-1\rightarrow n$, for each labeled node $i\in V(G)=\{1,...,n-1\}$, the probability of attachment is 
\eq{
p_i=\frac{1}{n-1}.
}
The RRT has been studied for many years \cite{wikstrom2020history} and in great mathematical depth \cite{zhang2015number}. It has degree distribution approaching $p(k)=2^{-k}$ \cite{janson2005asymptotic}.

\subsection{Barab\'asi-Albert model}\label{sssec:ba}

The BA tree \cite{barabasi1999emergence} has growth rule
\eq{
P_{n}(G'|G)=\frac{k_{v(G,G')}(G)}{\sum_{i=1}^{n-1}k_i(G)},
}
where $G$ is an $(n-1)$-node labeled tree and $G'\in\mathcal{G}(G)$.  In more simple notation indexed by node $i\in\{1,...,n-1\}$, 
\eq{
p_i=\frac{k_i(G)}{2(n-2)},
}
where we have used $\sum_ik_i(G)=2m(G)=2(n-2)$. It has powerlaw-tailed degree distribution of the form 
\eq{
p(k)&=\frac{4}{k(k+1)(k+2)}\\
&\sim k^{-3}.
}
Numerous extensions and applications have been considered \cite{dorogovtsev2000structure}.

\subsection{Linear preferential attachment with additive shift}\label{sssec:lpa}

Linear preferential attachment with additive shift $\delta$ has attachment kernel is $f(G,G')=k_{v(G,G')}(G)+\delta$. In the labeled case, the denominator of the growth rule distribution for $n-1\rightarrow n$ can be evaluated, as
\eq{
\sum_{G'\in\mathcal{G}(G)}f(G,G')&=\sum_{i=1}^{n-1}(k_i(G)+\delta)\\
&=(2+\delta)n-(4+\delta).
}
having used $\sum_ik_i(G)=2m(G)=2(n-2)$. This implies the probability of attaching to node $i\in V(G)=\{1,...,n-1\}$ in growth step $n-1\rightarrow n$ is given by
\eq{
p_i=\frac{k_{i}(G)+\delta}{(2+\delta)n-(4+\delta)},
}
with $\delta\in[-1,\infty)$. This model is also known as affine preferential attachment \cite{marchand2020influence}, or preferential attachment with initial attractiveness \cite{dorogovtsev2000structure}. The degree distribution has the asymptotic form
\eq{
p(k)\sim k^{-\gamma},
}
with
\eq{
\gamma=3+\delta,
}
and thus has a diverging second moment for any $\delta\le 0$. At $\delta=0$ the BA model and $\gamma=3$ are recovered.

\section{Powerlaw exponents in $\mathrm{UPA}(\delta)$ and $\mathrm{UPA}$}
\label{app:exponent_finding}

In this section we obtain implicit expressions for 
\begin{itemize}
\item[(i)] the scaling exponent $\xi$ of $\mathrm{UPA}$, 
\item[(ii)] the scaling exponent $\xi(\delta)$ of $\mathrm{UPA}(\delta)$ in the anomalous (subextensive) scaling regime $\delta\in(-1,\delta^*)$, and 
\item[(iii)] the asymptotic leaf-fraction $c(\delta)$ of $\mathrm{UPA}(\delta)$ in the linear scaling regime $\delta\in(\delta^*,\infty)$. 
\end{itemize}
We also find $\delta^*$. We show that $\xi(\delta^*)=1$ (indicating linear-order scaling) and $c(\delta^*)=1$ (indicating leaf-proliferation).

In what follows we rely on the following theorem from Ref.~\cite{doi:10.1137/1009002}, pertaining to the {\it minimal solution} of 2nd order linear recursions. If $x_{t+1}+a_tx_{t}+b_tx_{t-1}=0$ for all $t\ge 2$, then if there exists a minimal solution ($x_\infty=0$), the following consistency relation among $x_1$ and $x_2$ must hold:
\eq{
\label{eq:recursion_theorem}
\frac{x_2}{x_1}=\frac{-b_2}{a_2-\cfrac{b_3}{a_3-\cfrac{b_4}{a_4-\cdots}}}.
}

Since we have $m_0,m_1,m_2$ in terms of external parameters such as $c,\delta,\xi$, and since we have a 2nd order recursion applying at all $\ell\ge 2$, the above theorem applies in each case of interest herein. In particular, it provides another constraint on said parameters $c,\delta,\xi$, by requiring that the initial data $m_1$ and $m_2$ be consistent in the above sense. In each case, it leads to a unique positive solution; $\xi$ for $\mathrm{PA}$, $\xi(\delta)$ for $\mathrm{PA}(\delta<\delta^*)$, and $c(\delta)$ for $\mathrm{PA}(\delta>\delta^*)$.

\subsection{Powerlaw exponents in $\mathrm{UPA}(\delta)$ for $\delta<\delta^\ast$ and $\mathrm{UPA}$}
\label{ssec:upa_sublinear}

Consider scaling ansatz $L=n-bn^\xi$, $M_\ell=w_\ell n^\xi$, with $b=\sum_{\ell}w_\ell$, also defining $r_\ell=w_\ell/b$. The parameter $\xi$ is only assumed to be in $(0,1)$. Approximating to first order in $n^{\xi-1}$ the $\mathrm{UPA}(\delta)$ equations become
\eq{
\xi & =(1+\delta)(1-r_0),\\
\xi r_0 &= (1+\delta)(r_1-r_0),\\
\xi r_1 &= (1+\delta)(1+r_2-2r_1)-r_1,\\
\xi r_{\ell } &= (1+\delta)r_{\ell + 1}-(2+2\delta+\ell )r_{\ell }+(\ell+\delta)r_{\ell - 1}, \  \ell\ge 2,
}
with the special case of $\delta=0$ recovering unlabeled PA.

Thus, in terms of $\xi$, at $\ell=0$ we have
\eq{
r_0 = 1-\frac{\xi}{1+\delta};
}
at $\ell=1$,
\eq{
r_1&=1-\left(\frac{\xi}{1+\delta}\right)^2,
}
and at $\ell=2$, 
\eq{
r_2= \frac{3+2\delta+\xi}{1+\delta} \left[1-\left(\frac{\xi}{1+\delta}\right)^2\right]-1.
}
For all $\ell\ge 2$, the following form of recursion holds:
\eq{
r_{\ell+1}+a_\ell r_{\ell}+b_\ell r_{\ell-1}=0,
}
where $a_\ell$ and $b_\ell$ are linear functions of $\ell$, with dependence on $\delta$ and $\xi$. The coefficients can be read off from
\eq{
r_{\ell + 1}-\frac{\xi+2+2\delta+\ell }{1+\delta}r_{\ell }+\frac{\ell+\delta}{1+\delta}r_{\ell - 1}=0, \  \ell\ge 2,
}
i.e.,
\eq{
a_\ell&=-\frac{\xi+2+2\delta+\ell}{1+\delta},\\
b_\ell&=\frac{\ell+\delta}{1+\delta},
}
We impose the requirement of a decaying solution by demanding $r_\infty=0$. Applying Eq.~\eqref{eq:recursion_theorem}, a minimal solution will satisfy
\eq{
\frac{r_{2}}{r_1}=\frac{-b_2}{a_2-\cfrac{b_3}{a_3-\cfrac{b_4}{a_4-\cdots}}}.
}
The LHS is
\eq{
\frac{r_2}{r_1}&=\left[\frac{3+2\delta}{1+\delta}+\frac{\xi}{1+\delta}\right]-\left[1-\left(\frac{\xi}{1+\delta}\right)^2\right]^{-1}\\
&= \frac{3+2\delta+\xi}{1+\delta}-\frac{(1+\delta)^2}{(1+\delta)^2-\xi^2}.\\
}
\begin{widetext}
By being a continued fraction involving ratios of $a$s to $b$s, a factor $-1/(\delta+1)$ common to all terms cancels. As such, we arrive at the final implicit expression for $\xi(\delta)$:

\eq{
\frac{3+2\delta+\xi}{1+\delta}-\frac{(1+\delta)^2}{(1+\delta)^2-\xi^2} &
=\frac{2+\delta}{\xi+4+2\delta+\cfrac{3+\delta}{\xi+5+2\delta+\cfrac{4+\delta}{\xi+6+2\delta+\cdots}}}.
}

Note that since powerlaw exponent is $\gamma(\delta)=1+\xi(\delta)$, we can also state the above in terms of $\gamma(\delta)$ and $\delta$:
\eq{
\frac{2+2\delta+\gamma}{1+\delta}-\frac{(1+\delta)^2}{(\gamma+\delta)(2+\delta-\gamma )} =\frac{2+\delta}{\gamma+3+2\delta+\cfrac{3+\delta}{\gamma+4+2\delta+\cfrac{4+\delta}{\gamma+5+2\delta+\cdots}}}.
}
In the special case of $\xi=1$ (emergence of giant nucleus), we have $\delta=\delta^*$ defined as the solution in $[0,1]$ to
\eq{
\label{eq:sublin_delta1}
\frac{4+2\delta}{1+\delta}-\frac{(1+\delta)^2}{(1+\delta)^2-1}&=\frac{\delta+2}{5+2\delta+\cfrac{\delta+3}{6+2\delta+\cfrac{\delta+4}{7+2\delta+\cdots}}}.
}
\end{widetext}
This is also the relation obtained by setting $c=1$ in the implicit equation for $c(\delta)$ derived below in the linear scaling regime $\delta>\delta^*$. The numerical solution to Eq.~\eqref{eq:sublin_delta1} is $\delta^*=0.19831...$; the leaf-proliferation transition therefore occurs at {\it positive} $\delta^*$, implying that $\mathrm{UPA}=\mathrm{UPA}(0)$ is indeed leaf-proliferating, and hence fits within the current scenario of $\mathrm{UPA}(\delta)$ for $\delta<\delta^*$. In particular, when we set $\delta=0$, the relation simplifies to
\eq{
3+\xi-\frac{1}{1-\xi^2}=\frac{2}{\xi+4+\cfrac{3}{\xi+5+\cfrac{4}{\xi+6+\cdots}}},
}
which has numerical solution $\xi=0.84355...$ (implying UPA has powerlaw exponent $\gamma_0=1.84355...$). Since $\gamma_0=1+\xi$, we can also write
\eq{
2+\gamma_0-\frac{1}{(\gamma_0-1)\gamma_0}=\frac{2}{3+\gamma_0+\cfrac{3}{4+\gamma_0+\cfrac{4}{5+\gamma_0+\cdots}}}.
}

\subsection{Powerlaw exponent in $\mathrm{UPA}(\delta)$ for $\delta>\delta^*$}\label{app:upa_superlinear}

For $\delta>\delta^*$, numerical evidence suggests that linear scaling holds. We apply an ansatz akin to that for UUA: $L\simeq cn,M_\ell \simeq m_\ell n$, $\ell\ge 0$. In this case, the $\mathrm{UPA}(\delta)$ equations become
\eq{
\label{eq:system} 
q_\delta c &= (1+\delta)(1-c)+c,\\
q_\delta m_0 &= (1+\delta)(m_1-m_0),\\
q_\delta m_1 &= (1+\delta)m_2-(2+2\delta+1)m_1 +(1+\delta)(1-c)\\ 
q_\delta m_\ell &= (1+\delta)m_{\ell+1}-(2+2\delta+\ell)m_\ell+(\ell+\delta)m_{\ell-1},
}
with the latter equation holding for $\ell\ge 2$, and with
\eq{
q_\delta=(1+\delta)(2(1-c)-m_0)+c.
}
The first few equations yield $m_2(c),m_1(c),m_0(c),q_\delta(c)$. The 2nd order recursion holds for all higher $\ell$:
\eq{
m_{\ell+1}-\frac{2+2\delta+q_\delta+\ell}{1+\delta}m_\ell+\frac{\ell+\delta}{1+\delta}m_{\ell-1}=0, \ \ell\ge 2,
}
or, with 
\eq{
a_\ell&=-\frac{2+2\delta+q_\delta+\ell}{1+\delta},\\
b_\ell&=\frac{\ell+\delta}{1+\delta},
}
\eq{
m_{\ell+1}+a_\ell m_\ell+b_\ell m_{\ell-1}=0, \ \ell\ge 2.
}
We examine the minimal solution as per Ref.~\cite{doi:10.1137/1009002}, using Eq.~\eqref{eq:recursion_theorem} to obtain
\eq{
\frac{m_{2}}{m_{1}}=\frac{2+\delta }{4+2\delta+q_\delta +\cfrac{3+\delta}{5+2\delta+q_\delta+\cfrac{4+\delta}{6+2\delta+q_\delta+\cdots}}}.
}
We have from Eqs.~\eqref{eq:system} that
\eq{
m_0&=\left(2-\frac{1}{1+\delta}-\frac{1}{c}\right)(1-c),\\
m_1&=\left(\frac{1}{1+\delta}+\frac{1}{c}\right)m_0,\\ 
m_2&=\left(1+\frac{2}{1+\delta}+\frac{1}{c}\right)m_1 -(1-c).
}
\begin{widetext}
Hence,
\eq{
\frac{m_2}{m_1}&=\left(1+\frac{2}{1+\delta}+\frac{1}{c}\right)-\left(\frac{1}{1+\delta}+\frac{1}{c}\right)^{-1}\left(2-\frac{1}{1+\delta}-\frac{1}{c}\right)^{-1}.\\
}
Our implicit equation for $c(\delta)$ is thus
\eq{
\left(1+\frac{2}{1+\delta}+\frac{1}{c}\right)-\left(\frac{1}{1+\delta}+\frac{1}{c}\right)^{-1}\left(2-\frac{1}{1+\delta}-\frac{1}{c}\right)^{-1}=\frac{2+\delta  }{4+2\delta+q_\delta +\cfrac{3+\delta}{5+2\delta+q_\delta +\cfrac{4+\delta }{6+2\delta+q_\delta+\cdots}}},
}
\end{widetext}
with $q_\delta=1+(1+\delta)(1-c)/c$. The condition $c=1$ is achieved at $\delta=\delta^*$; then $q_\delta=1$ and the equation becomes identical to Eq.~\eqref{eq:sublin_delta1} obtained in Sec.~\ref{ssec:upa_sublinear} under condition $\xi(\delta^*)=1$.

\section{Simulation methodology}\label{app:simulations}

To simulate unlabeled growth, we need to construct distributions over the orbit to which new nodes attach. This requires the computation of orbit partitions, and their re-computation at each timestep. Once orbits are computed, degree-values of representatives are obtained, from which the distribution $p_u=\phi(G,u)/\sum_{u'\in D(G)}\phi(G,u')$ is constructed. To distinguish orbits we utilize {\it complete node invariants}, i.e., properties taking a unique value depending only on node orbit.

\subsection{Orbit finding}\label{app:orbit_finding}
Here we describe an algorithm to extract all automorphism orbits of a labeled tree, providing a grouping according to unlabeled structural position. To identify orbits, we compute a {\it complete node invariant} $\phi(G,i)$ at each labeled node $i$. By definition, it has a distinct values $\phi(G,i)\ne\phi(G,i)$ for all pairs $i,j$ not sharing an orbit. This defines {\it complete}, which is in addition to the requirement of having a common value $\phi(G,i)=\phi(G,j)$ for all pairs $i,j$ in a common orbit (as for all node invariants). The computation of every node's complete node invariant provides an automorphism orbit decomposition, after a grouping by orbit class via direct comparison of the computed $\phi(G,i)$-values.

We choose the {\it lexicographic representation} as a node invariant. The lexicographic representation of graph $G$ by node $r\in V(G)$ is a recursively constructed encoding of $G$ from $r$'s vantage point. It is straightforwardly computable: root $G$ with $r$, and build its string $s_r(G,r)$ by a sorted concatenation of $s_{r}(G,j)$ for all children $j\sim r$; recurse out to leaves which are assigned a base symbol, by convention a pair of parentheses ``$()$''. The symbol $s_r(G,i)$ at node $i$ is built after symbols from all children $j$ of $i$ (relative to root $r$); $\{s_r(G,j)\}_j$ are obtained, sorted, and concatenated into $s_r(G,i)=(s_r(G,{j_1})\cdots s_{r}(G,j_{k}))$, with $k=\mathrm{deg} \ i - 1$ . When the recursion completes, the string passed to root $r$ is a complete node invariant $\phi(G,r)$: it uniquely encodes the full unlabeled tree structure from the vantage point of $r$. After computing $\phi(G,r)$ for all $r\in V(G)$, we compare elements of $\{\phi(G,r)\}_r$ to form node groupings by automorphism orbit. The final string of brackets, e.g., $(((()))()(()))$, may be arbitrarily encoded to store the value of $\phi(G,r)$. E.g., as just a string of parentheses, or as a binary string $00001110100111$, etc.--- as long as the mapping is one-to-one, complete node invariance is retained.

\section{Growing non-tree models}\label{app:non_tree}

The models studied herein have been of growing trees, but numerous labeled growing network models are not trees. For instance, a new node may arrive and attach to a fixed or random number $m>1$ of the existing nodes. In this setting, we no longer can associate unlabeled growth outcomes with individual orbits.

Assume an arbitrary initial graph $G_0$ of size $n_0=m$. The set of unlabeled growth outcomes from an $(n-1)$-node graph is the set of isomorphism classes arrived at by a new node with $m$ link-attachments into $G$. Let the set of orbits of $G$ be denoted $u\in D(G)$. For those orbits with any degeneracy, i.e., $u:|u|>1$, then up to $|u|$ of those representatives may be attached to; due to their indistinguishability, the only pertinent information is how many attachments were allocated into $u$. Thus overall, the growth rule assigns probability to each of the ways in which $m$ links can be placed into $|D(G)|$ distinguishable boxes of sizes $\{|u|\}_{u\in D(G)}$. Finally, and most nontrivially, we must account for any symmetries arising as a consequence of differing combinations of such attachments, checking for isomorphism among those constructed attachment possibilities. The resulting set $\mathcal{G}(G)$ of graphs growable from $G$ is the sample space for the probability distribution $P_n(G'|G)$ specifying growth rules.

This procedure allows construction of the possible attachment rules with $m>1$. Namely, as a function of the $m$ orbits attached to, $\{u_i\}_{i=1}^m$ (some of which may be duplicates), we may examine a given preference function $\phi(G,\{u_i\}_{i=1}^m)\ge 0$. For instance, with proportionality to $m$ factors of the form $k_{u_i}+\delta$, generalizing $\mathrm{UPA}(\delta)$.

Labeled linear preferential attachment has been shown to be asymmetric w.h.p. for $m\ge 3$ \cite{luczak2019compression}. As such, we anticipate many cases of equivalence between the labeled and unlabeled growth models for $m\ge 3$. Having shown cases of significant nonequivalence at $m=1$ in this work, a notable case of interest is $m=2$.

\end{document}